\documentclass[useAMS,usenatbib]{mnras}

\usepackage{graphicx}
\usepackage{amsmath}
\usepackage{amssymb}
\usepackage{multicol}
\usepackage{bm}
\usepackage{pdflscape}
\usepackage{float}
\usepackage{tipa}
\usepackage{amsmath}
\usepackage{breqn}
\usepackage{subfig}
\usepackage{color,soul}

\hypersetup{final}

\title[]{Inhibition of the electron cyclotron maser instability in the dense magnetosphere of a hot Jupiter}
\author[]{Daley-Yates S., Stevens I. R.}

\author[S. Daley-Yates, I. R. Stevens]{S. Daley-Yates$^{1}$\thanks{E-mail:
sdaley@star.sr.bham.ac.uk}, I. R. Stevens$^{1}$\\
$^{1}$School of Physics and Astronomy, University of Birmingham, Edgbaston, Birmingham, B15 2TT\\}

\begin{document}

\date{}

\pagerange{\pageref{firstpage}--\pageref{lastpage}} \pubyear{2018}

\maketitle

\label{firstpage}

\begin{abstract}

Hot Jupiter (HJ) type exoplanets are expected to produce strong radio emission in the MHz range via the Electron Cyclotron Maser Instability (ECMI). To date, no repeatable detections have been made. To explain the absence of observational results, we conduct 3D adaptive mess refinement (AMR) magnetohydrodynamic (MHD) simulations of the magnetic interactions between a solar type star and HJ using the publicly available code PLUTO. The results are used to calculate the efficiency of the ECMI at producing detectable radio emission from the planets magnetosphere. We also calculate the frequency of the ECMI emission, providing an upper and lower bounds, placing it at the limits of detectability due to Earth's ionospheric cutoff of $\sim 10 \ \mathrm{MHz}$. The incident kinetic and magnetic power available to the ECMI is also determined and a flux of $0.069 \ \mathrm{mJy}$ for an observer at $10 \ \mathrm{pc}$ is calculated. The magnetosphere is also characterized and an analysis of the bow shock which forms upstream of the planet is conducted. This shock corresponds to the thin shell model for a colliding wind system. A result consistent with a colliding wind system. The simulation results show that the ECMI process is completely inhibited by the planets expanding atmosphere, due to absorption of UV radiation form the host star. The density, velocity, temperature and magnetic field of the planetary wind are found to result in a magnetosphere where the plasma frequency is raised above that due to the ECMI process making the planet undetectable at radio MHz frequencies.

\end{abstract}

\begin{keywords}
planet-star interactions - planets and satellites: aurorae, magnetic fields  - radio continuum: planetary systems.
\end{keywords}

\section{Introduction} 
\label{sec:intro}

Understanding the evolution of giant exoplanets is important as they give us insights into the long term evolution of their host planetary systems. The reason for this is that giant planet formation is inhibited close to the parent star \citep{Murray-Clay2009}, and they must undergo migration across many orders of magnitude of radius to be found in short period orbits ($<$ 10 days). Such planetary migration has implications for the evolution of their entire host system \citep{Fogg2005, Petrovich2016, Alvarado-Montes2017}.

Giant planets in such close orbits are know as hot Jupiters (HJs). As a consequence of their close orbits, the planets atmosphere receives large amounts of UV radiation, leading to atmospheric blow off or the establishment of a hydrodynamic planetary wind \citep{Matsakos2015} which can reach temperatures of $10^{4}$ K  \citep{Shaikhislamov2014}.

Recent studies have monitored increased emission in X-ray as well as metal lines of HJ hosting stars \citep{Shkolnik2008, Pillitteri2015, Miller2015}. \cite{Gurdemir2012} conclude that enhanced emission in Ca II H and K lines of HD 179949 is due to planetary material accreting onto the stellar surface, inducing enhanced chromospheric emission. 

Many of the dynamic behaviour and observable signatures of extrasolar planetary systems can be characterised through the paradigm of Star-Planet Interaction (SPI). Traditional planet detection techniques including the transit and radial velocity methods (see \cite{Wright2012} for a review of the various methods) rely centrally on observing the interplay between the planet and the host star. At many wavelengths such as infra-red or visible, the planet is simply too dim compared to the luminosity of its host star, making direct detection rare.

Radio wavelengths provide an alternative method for exoplanet observation to those described above, with the possibility of direct planet detection \citep{Stevens2005, Zarka2007}. Planet hosting stars are typically radio quiet, therefore SPI processes that result in significant, detectable radio emission will be observed despite the presence of the host star. Star-Planet Magnetic Interaction (SPMI) is a prime candidate for generating such observable radio emission. Examples of this type of interaction include magnetic reconnection between polar magnetic field lines of the star and planet as well as reconnection in the magneto tail of the planet. Another mechanism for SPMI is the amplification of the Electron Cyclotron Maser Instability (ECMI). Emission via this mechanism is due to incident solar wind power, both ram and magnetic, on the planetary magnetic field. The efficiency of the ECMI for producing radio emission is the main topic of this paper.

\subsection{Electron Cyclotron Maser Instability: A Means for Detecting Hot Jupiters}

ECMI is a process where electrons emit radiation due to confinement by magnetic field lines along which they travel, gyrate and accelerate around \citep{Melrose1982, Dulk1985}. The emission is almost 100 \% circularly polarised and directionally beamed. The environment around exoplanets such as HJs is thought to be well suited to producing this form of emission, due to a high magnetic field strengths and strong incident stellar wind power.

Many studies have investigated the conversion of incident stellar wind power to ECMI emission \citep{Kivelson1997, Stevens2005, Zarka2007, Hess2011}. Conversion efficiencies up to 0.2 \% of the incident stellar wind power has been proposed by \cite{Zarka2007} as a consequence of a \textit{radio-magnetic Bode's Law} (see the aforementioned paper for details). Such high power conversion rates would lead to emission of the order of 1 - 10 mJy of detectable emission (from a candidate system at 10 pc) \citep{Stevens2005}. Such high fluxes indicate that, for a full understanding of the observable radio signatures of SPMI, the host star must be considered explicitly in any model as it dynamic influences the evolution and observable properties of short period HJs. 

ECMI emission is also seen at radio wavelengths from a number of other stellar sources. Examples include:
\begin{itemize}
\item Low-mass dwarf stars, sometimes referred to as Ultracool Dwarfs (UCDs). The observed radio emission is characterised by bright, circularly polarised bursts at GHz frequencies that are of short duration (1-100 minutes). \cite{Williams2017} provides a summary of recent results on this class of object.
\item Magnetic early type stars can also show highly polarised, rotationally modulated bursts, consistent with ECMI emission \cite{Trigilio2011}. See also \cite{Das2017} for results on the magnetic Bp star HD 1333880 (HR~Lup) and the main sequence  B2.5V star HR 5907 \citep{Leto2018}.
\item RS~CVn type systems, which can also show highly circularly polarised burst like emission in the binary system. See \cite{Slee2008} for observations of the RS CVn system HR~1099.
\end{itemize}

\subsection{Radio Emission from Hot Jupiter Exoplanets}

Infra-red, visible and UV emission from exoplanets is difficult to detect from Earth due to the high luminosity of the host star \citep{Zarka2007}. Main sequence solar type stars are however quiet in the low frequency radio spectrum, making the detection of radio emission from the planet feasible \citep{Griesmeier2007}.

As discussed in the previous section, radio ECMI emission is directly related to the magnetic field strength of the planet. Therefore observations of this type of emission can provide an indirect means of measuring and classifying the planetary magnetic field. This quantification of the magnetic filed gives us constraints on internal structure models and planetary rotation, informing us about the evolutionary history of the planet and host system \citep{Hess2011}.

HJs present the best opportunity for studding extra solar planetary radio emission. This due to their close proximity to their host stars and their potentially strong magnetic fields.

The SPMI origin of non-thermal radio emission is by nature transient. If the planet orbits within the Alfv\'{e}n surface of the star, reconnection events between the stellar and planetary magnetospheres will occur \citep{Strugarek2014a}. This will depend on the dynamic behaviour of the stellar surface. In addition, the amplification of ECMI emission by the incident stellar wind is inherently dependent upon the wind conditions along the trajectory of material from the stellar surface to the planet. transient stellar surface events such as coronal mass ejections (CME) and flares will modify the ram pressure and magnetic energy density of the wind, leading to higher intensity EMCI emission \citep{Llama2013, Vidotto2015}. The transient nature of these phenomenon make detection challenging.

A recent study by \cite{Weber2017} explored analytically the ECMI process in the magnetosphere of HJs. The study concludes that enhanced particle density, due to the planetary wind, inhibits the ECMI process for magnetic field strength $< 50 \mathrm{G}$. Our study aims to build on this result by conducting self consistent numerical MHD simulations of a HJ system, including the host star. Simulations produce dynamic features which are not captured in analytic calculations. 

\subsection{Analogies in the Solar System}

Justification for the expectation of bright ECMI emission is found via analogy to the solar system planets and the Jovian moons. For example, the Jovian auroral radio emission, at MHz frequencies is dominated by ECMI processes from keV electrons in the auroral regions of the planet \citep{George2007}. At higher frequencies ($\sim$~GHz) the emission from electrons in the planets radiation belts dominates, however at orders of magnitude lower intensity than EMCI emission \citep{Zarka2007}. There is a cut off at lower frequencies due to the Earth's ionosphere (see Section \ref{sec:detect}). ECMI emission has also been detected from Saturn, Uranus and Neptune by space based missions \citep{Zarka1998}. 

\cite{Zarka2007} theoretically investigated HJ radio emission via  extrapolation from the solar system planets and the Jovian moons. By considering the incident solar wind power and the effective obstacle area of each body (magnetispheric radius), the emitted radio power can be estimated. Using this approach, \cite{Zarka2007} developed a Generalised "Radio Bode Law" where there is almost a one to one relationship between the incident power and the emitted radio power, given an efficiency factor of $\sim~2 \times 10^{-3}$. This leads to an estimate of emitted radio power for a HJ in the region of $10^{14}$~--~$10^{16}$ W (see Fig. 6 of \cite{Zarka2007}). Radio power at this magnitude results in detectable emission of the order of a few mJy \citep{Stevens2005}. This flux is easily detectable by the current generation of telescopes.  

\subsection{Prospects for Detectability}
\label{sec:detect}

The possible detection of exoplanet radio emission is hampered by the presence of the Earths ionosphere. The ionosphere is a partially ionised layer of the atmosphere, with a typical altitude of 50-1000 km. It is very dynamic, with the electron density varying dramatically as a function of space and time, being mainly affected by UV and X-ray solar emission and by charged particles from the solar wind. The ionospheric plasma frequency (see Section \ref{sec:ECMI}) typically ranges between 1-10~MHz, but sometimes (e.g during sporadic E-layer events) can reach as high as 200~MHz. Cosmic radio waves with a frequency below the ionospheric plasma frequency will be reflected by the ionosphere and thus not reach an Earth-bound telescope. Understanding and accounting for the ionosphere is a major concern for low frequency radio telescopes (operating at $<$ 300~MHz), such as the GMRT, LOFAR, LWA and MWA and SKA-Low in future \citep{Intema2009}. 

Thus far, there have been several observational studies which have attempted to detect exoplanet radio emission including LOFAR/VLT observation by \cite{Knapp2016} and LOFAR observations by \cite{Turner2017}. Both report no detections. \cite{Murphy2014} have placed limits on low-frequency radio emission from 17 known exoplanetary systems with the Murchison Widefield Array. They detected no radio emission at 154 MHz, and put 3$\sigma$ upper limits in the range $15.2 - 112.5$ mJy. The only reported detection thus far is of 150 MHz emission from HAT-P-11b by \citep{Etangs2013} with a flux of 3.87 mJy. This signal however was not detected by repeat observations and is thus not conclusive.

This study aims to further investigate this conclusion by performing high resolution simulations of both the exoplanetary magnetosphere and global evolution of the stellar wind in which the exoplanet is embedded. this is done using magnetohydrodynamic, adaptive mesh refinement numerical simulations. The following section will detail the governing conservation equations and the specific setup used to simulate the stellar and exoplanetary bodies and the mechanism and conditions needed for ECMI emission to be produced.

\subsection{Simulations}
\label{sec:sim}

Numerous numerical studies have been carried out in recent years which have simulated many aspects of SPI, examples range from detailed studies of the close in atmospheres of HJs \citep{Strugarek2014a, Khodachenko2015, Vidotto2015}, the atmospheres of their host stars \citep{Alvarado-Gomez2016, Fares2017}, to the global structure of the planetary-stellar wind interaction \citep{Bourrier2013, Bourrier2016, Owen2014, Alexander2015}. The study presented here builds on this work, specifically on the simulations conducted by \cite{Matsakos2015}, extending it to investigate ECMI emission in the context of SPMI. 

\section{Modelling}
\label{sec:modelling}

The models constructed here follow the approach used by \cite{Matsakos2015}, however, instead of using a static numerical grid unwhich the MHD equations are solved, we employ the method of Adaptive Mesh Refinement (AMR). The advantage of using a static mesh is that the resolution, and therefore the memory usage is kept constant throughout the simulation. This is not the case with AMR, whose advantage of is that evolution of material not in the vicinity of either the star or the planet, can be actively traced. This allows for a high resolution study of the large scale structure of material evaporated from the planet to be studied. This is an important consideration if observable effects such as asymmetric transit depths \citep{Llama2013} are to be investigated.

\cite{Chadney2015} argue that the atmosphere of a HJ is either in a hydrostatic or hydrodynamic state depending on it's distance from it's host star. The transition between these to regimes is an orbital distance of $\sim 0.5 \ \mathrm{au}$, for a solar type star. Since the orbital distance in this work is set to $0.047 \ \mathrm{au}$, we make the assumption that the planetary atmosphere is purely hydrodynamic in nature. This assumption is further reinforced by work carried out by \cite{Murray-Clay2009} who model heating and cooling, ionization balance, tidal gravity, and pressure confinement by the host star wind when studying the nature of the mass-loss from UV evaporated HJs. They found that the resulting planetary wind takes the form of a Parker wind emitted from the planets day side. For this reason, the both the stellar and planetary winds in this study take the form of that described by \cite{Parker1958} and evolve according to the equations of magnetohydrodynamics (MHD). The following section details these equations.

\subsection{Magnetohydrodynamics}

The equations of magnetohydrodynamics are solved in a frame co-rotating with the orbital motion of the planet. These equations are:
\begin{equation}
  \label{eq:mass}
  \frac{\partial \rho}{\partial t} + \bm{\nabla} \cdot \left( \rho \bm{v} \right) = 0
\end{equation}
\begin{equation}
  \label{eq:momentum}
  \frac{\partial \bm{v}}{\partial t} + \left( \bm{v} \cdot \bm{\nabla} \right) \bm{v}  
  + \frac{1}{4 \pi \rho} \bm{B} \times  \left( \nabla \times \bm{B} \right) 
  + \frac{1}{\rho} \nabla p = \bm{g} + \bm{F}_{\mathrm{co}}
\end{equation}
\begin{equation}
  \label{eq:energy}
  \frac{\partial p}{\partial t} + \bm{v} \cdot \bm{\nabla} p  
  + \gamma p \nabla \cdot \bm{v} = 0
\end{equation}
\begin{equation}
  \label{eq:magnetic}
   \frac{\partial \bm{B}}{\partial t} 
   + \nabla \times \left( \bm{B} \times \bm{v} \right) = 0.
\end{equation}
Where $\rho$, $\bm{v}$, $\bm{B}$, $p$, $\bm{g}$ and $\bm{F}_{\mathrm{co}}$ are, respectively, the density, velocity, magnetic field, pressure, acceleration due to gravity and acceleration due to the co-moving frame. $\bm{F}_{\mathrm{co}}$ is the sum of the centrifugal and Coriolis forces: $\bm{F}_{\mathrm{co}}~=~\bm{F}_{\mathrm{centrifugal}}~+~\bm{F}_{\mathrm{coriolus}}$ which are given by
\begin{equation}
  \bm{F}_{\mathrm{centrifugal}} = 
  - \left[ \bm{\Omega}_{\mathrm{fr}} \times \left( \bm{\Omega}_{\mathrm{fr}} 
  \times \bm{R} \right) \right] = \bm{\Omega}_{\mathrm{fr}}^{2} 
  \left( x \hat{x} + y \hat{y} \right)
\end{equation}
and
\begin{equation}
  \bm{F}_{\mathrm{coriolus}} = 
  - 2 \left( \bm{\Omega}_{\mathrm{fr}} \times \bm{v} \right) 
  = 2 \Omega_{\mathrm{fr}} \left( v_{x} \hat{x} + v_{y} \hat{y} \right),
\end{equation}
where $\bm{\Omega}_{\mathrm{fr}}$ is the angular frequency of the frame in which the calculations are being conducted, $R$ is the radial distance from the origin.

An adiabatic equation of state is used to close the MHD equations. Both the stellar and planetary winds are assumed to be isothermal, to approximate this behaviour, $\gamma = 1.05$ in equation (\ref{eq:energy}). 

In the following sections the equations which govern the stellar and planetary winds as well as the their interiors, magnetic fields and orbital parameters are presented.

\subsection{Stellar and Planetary Models}
\label{sec:models}

Since both the stellar and planetary winds and magnetic fields are initialised using the exact same equations, they are presented here in their most basic form. When distinguishing between equations which are specific to, or contributions specifically from one of the individual bodies, a subscript of either $_\ast$ or $_\circ$ is used to indicate that it applies to the star or planet respectively. For example, when referring to the radial distance, there is the radius from the stellar centre $r_{\ast}~=~\sqrt{x_{\ast}^2 + y_{\ast}^2 + z_{\ast}^2}$ and the radius from the planet centre $r_{\circ}~=~\sqrt{(x_{\ast} - a)^2 + y_{\ast}^2 + z_{\ast}^2}$, where $a$ is the orbital separation (in our models the star in centred on the origin such that $r_{\ast}~=~r$). Hence, when referring to the radius, $r$, it is assumed that the reader will understand that $r_{\ast}$ is implied when in the context of the star and  $r_{\circ}$ in the context of the planet. Therefore when an equation is comprised solely of variables relating to the star or planet, the subscripts are dropped.

\subsubsection{Wind Model and Initial Conditions} 
\label{sec:model_init}

The models for both the stellar and planetary winds are initialised according to the isothermal Parker wind model \citep{Parker1958}. The governing equation is \begin{equation}
\label{eq:parker_eq}
	\psi - \ln{\left( \psi \right)} = - 3 - 4 \ln{ \left( \frac{\lambda}{2} \right)} 
	+ 4 \ln{ \left( \xi \right) } + 2 \frac{ \lambda}{\xi}
\end{equation}
where the three dimensionless parameters $\psi$, $\lambda$ and $\xi$ are defined as:
\begin{equation}
   \psi \equiv \left( \frac{v^{\mathrm{init}}_{\mathrm{W}} (r) }{c_{\mathrm{s}}} \right)^2
\end{equation}
\begin{equation}
   \lambda \equiv \frac{1}{2} \left( \frac{v_{esc}}{c_{\mathrm{s}}} \right)^2
\end{equation}
\begin{equation}
   \xi \equiv \frac{r}{R}.
\end{equation}
$v^{\mathrm{init}}_{\mathrm{W}}(r)$, is the initial radial wind velocity, at the start of the simulation. $v_{esc}~=~\sqrt{2 G M/R}$ is the escape velocity and the isothermal sound speed is given by $c_{\mathrm{s}}~=~\sqrt{k_{\mathrm{B}} T / \mu_{\mathrm{mol}} m_{\mathrm{p}}}$, with $k_{\mathrm{B}}$ the Boltzmann constant, $T$ is the temperature, the mean molecular weight is $\mu_{\mathrm{mol}} = 0.5$, and $m_{\mathrm{p}}$ the proton mass. $\xi$ gives the distance from centre of the body, in units of the bodies radii. $R$ is the radius of either body. See Section \ref{sec:Circum} Fig. \ref{fig:parker_res} for the analytic and simulation result for equation (\ref{eq:parker_eq}).

Equation (\ref{eq:parker_eq}) is transcendental and must be solved numerically. Using an appropriate root finding algorithm, such as the Newton-Raphson method allows equation (\ref{eq:parker_eq}) to be solved for $v^{\mathrm{init}}_{\mathrm{W}}(r)$ for all values of $r$. The final step is to account for the rotation of the frame by assigning non-radial components to the wind velocity. This is done differently for the star than for the planet, as the star is located at the rotational axis of the frame. In the case of the star, the components are
\begin{multline}
\label{eq:vel_start}
   v_{\ast x}^{\mathrm{init}}= \sin(\theta_{\ast}) \left[ \right. \cos(\phi_{\ast}) 
   v^{\mathrm{init}}_{\ast \mathrm{W}}(r_{\ast}) 
   \\ + \sin(\phi_{\ast}) r_{\ast} \left( \Omega_{\mathrm{fr}} + \Omega_{\ast} \right) \left. \right], 
\end{multline}
\begin{multline}
   v_{\ast y}^{\mathrm{init}} = \sin(\theta_{\ast}) \left[ \right. \sin(\phi_{\ast}) 
   v^{\mathrm{init}}_{\ast \mathrm{W}}(r_{\ast}) 
   \\ + \sin(\phi_{\ast}) r_{\ast} \left( \Omega_{\mathrm{fr}} + \Omega_{\ast} \right) \left. \right], 
\end{multline}
\begin{equation}
   v_{\ast z}^{\mathrm{init}} = \cos(\theta_{\ast}) v^{\mathrm{init}}_{\ast \mathrm{W}}(r_{\ast}). 
\end{equation}
For the planet they are
\begin{multline}
   v_{\circ x}^{\mathrm{init}} = \sin(\theta_{\circ}) \left[ \right. \cos(\phi_{\circ}) 
   v^{\mathrm{init}}_{\circ \mathrm{W}}(r_{\circ})  
   \\ + \sin(\phi_{\ast}) r_{\ast} \Omega_{\mathrm{fr}} - a \Omega_{\mathrm{orb}} \sin({\cal M}) - \sin(\phi_{\circ}) r_{\circ} \Omega_{\circ} \left. \right], 
\end{multline}
\begin{multline}
   v_{\circ y}^{\mathrm{init}}  = \sin(\theta_{\circ}) \left[ \right. \sin(\phi_{\circ}) 
   v^{\mathrm{init}}_{\circ \mathrm{W}}(r_{\circ}) 
   \\ - \cos(\phi_{\ast}) r_{\ast} \Omega_{\mathrm{fr}} + a \Omega_{\mathrm{orb}} \cos({\cal M}) + \cos(\phi_{\circ}) r_{\circ} \Omega_{\circ} \left. \right], 
\end{multline}
\begin{equation}
   v_{\circ z}^{\mathrm{init}} = \cos(\theta_{\circ}) v^{\mathrm{init}}_{\circ \mathrm{W}}(r_{\circ}). 
\label{eq:vel_end}
\end{equation}
Where $r_{\ast, \circ}$, $\theta_{\ast, \circ}$ and $\phi_{\ast, \circ}$ are calculated from $x_{\ast, \circ}$, $y_{\ast, \circ}$ and $z_{\ast, \circ}$ by coordinate transformation. ${\cal M}$ is the mean anomaly, which accounts for the motion of the planet around the star. As the calculations are conducted in the frame of the orbiting planet, ${\cal M} = 0$ at all times. The parameters $\Omega_{\mathrm{orb}}$, $\Omega_{\ast}$ and $\Omega_{\circ}$ are the orbital, stellar and planetary rotational frequencies respectively. Our models assume that the planet is tidally bound to the star and that the star simply rotates at the same rate as the planetary orbit. This, together with the fact that the simulations are conducted in the rotating frame of the orbit, means that: $ \Omega_{\ast} = \Omega_{\circ} = \Omega_{\mathrm{orb}}$. with $\Omega_{\mathrm{orb}} = \sqrt{G M/R_{\mathrm{orb}}}$.

This frame work greatly simplifies the simulations, as the planets position is kept constant with time. However, as pointed out by \cite{Matsakos2015}, equations (\ref{eq:vel_start} - \ref{eq:vel_end}) are general and can accommodate the frame, star and planet having different rotational frequencies. To achieve this, Kepler's Equation should be solved to advance the position of the planet as a function of time (this would also allow for eccentric orbits). In addition, the model assumes that the centre of mass of the system is located at the centre of the star. If this assumption is relaxed, then the position of the star would also have to be updated in a similar manner. 

At all points in the wind, material is subject to acceleration due to gravity, this is applied as an acceleration vector according to
\begin{equation}
   \bm{g} (x,y,z) = -\frac{G M}{r^3} \bm{r},
\end{equation}
from both the star and the planet. The total gravitational acceleration is then the sum of the contributions from both bodies $\bm{g}_{\mathrm{tot}}(\bm{r}) = \bm{g}_{\ast}(\bm{r}_{\ast}) + \bm{g}_{\circ}(\bm{r}_{\circ})$. The initial pressure in the wind is found by solving 
\begin{equation}
\label{eq:pres_init}
  P^{\mathrm{init}} (x, y, z) = P  \exp{\left[ \lambda \left( \frac{1}{\xi} - 1 \right) - 
  \frac{\psi}{2} \right]}.
\end{equation}
$P$ is the pressure at the stellar or planetary surface, found by solving
\begin{equation}
   P = \frac{k_{\mathrm{B}} T \rho}{\mu_{\mathrm{mol}} m_{\mathrm{p}}},
\end{equation}
from the isothermal equation of state. Where $\rho$ and $T$ are the density and temperature at the stellar or planetary surface, given in Table \ref{tab:params}. The initial wind density profile is then
\begin{equation}
\label{eq:rho_init}
  \rho^{\mathrm{init}} (x, y, z) = \frac{\rho}{P} P^{\mathrm{init}} (x, y, z).
\end{equation}

\subsubsection{Magnetic Fields}

The magnetic field of both the star and planet are initialised as dipoles according to the following equation:
\begin{equation}
\label{eq:b_field}
    \bm{B}^{init} (x, y, z) = \frac{B_{eq} R^{3}}{r^{5}} 
    \left[ 3 x z \hat{x} + 3 y z \hat{y} + 
    \left( 3z^2 + r^2 \right) \hat{z} \right] 
\end{equation}
where $B_{eq}$ is the equatorial magnetic field of the body. The magnetic field external to the stellar and planetary surfaces is initialised as the sum of the stellar and planetary fields, $\bm{B}_{\mathrm{tot}}~=~\bm{B}_{\ast} + \bm{B}_{\circ}$. This ensures that the initial field has no discontinuities. This field is initially independent of the fluid dynamics. As the simulation evolves, the magnetic field reacts to and relaxes into the wind. This process leads to a magnetosphere that is largely dipolar in the inner regions. larger radii field lines are dragged with the flow forming an open magnetosphere. These open magnetic field lines correspond to a line of latitude on the surface of the bodies whose hight above the equator is a function of the strength of the magnetic field. A stronger field corresponds to a higher latitude (see Section \ref{sec:emission}).

\subsubsection{Stellar and Planetary Interiors}
\label{sec:Interiors}

The stellar and planetary surfaces are treated as internal boundaries to the simulation. Within the stellar or planetary surfaces, all quantities are held constant with time such that they are uniformly dens spheres. This simplifies the treatment of gravity within the bodies, allowing us to write the internal gravity as
\begin{equation}
  \bm{g}^{\mathrm{intern}} (x,y,z) = \frac{4}{3} \pi G \rho \bm{r}.
  \label{eq:grav_internal}
\end{equation}
From this, the hydrostatic condition leads to the following expression for the internal pressure;
\begin{equation}
  P^{\mathrm{intern}} (x,y,z)  = P + \frac{2}{3} \pi G \rho^{2} \left( R^{2} - r^{2} \right),
  \label{eq:pres_internal}
\end{equation}
where the the density, $\rho$, is given in Table \ref{tab:params}. 

Equations (\ref{eq:grav_internal} and \ref{eq:pres_internal}) are not strictly speaking consistent with the mass of the star or planet, as $M \ne \frac{4}{3} \pi G \rho R^{3}$ with $\rho$ the surface density and leads to a discontinuity in the gravitational field at the stellar and planetary surface. However, as this region is not in the computational active region, it will have zero influence on the solution.

\begin{figure}
\centering
\includegraphics[width=0.45\textwidth]{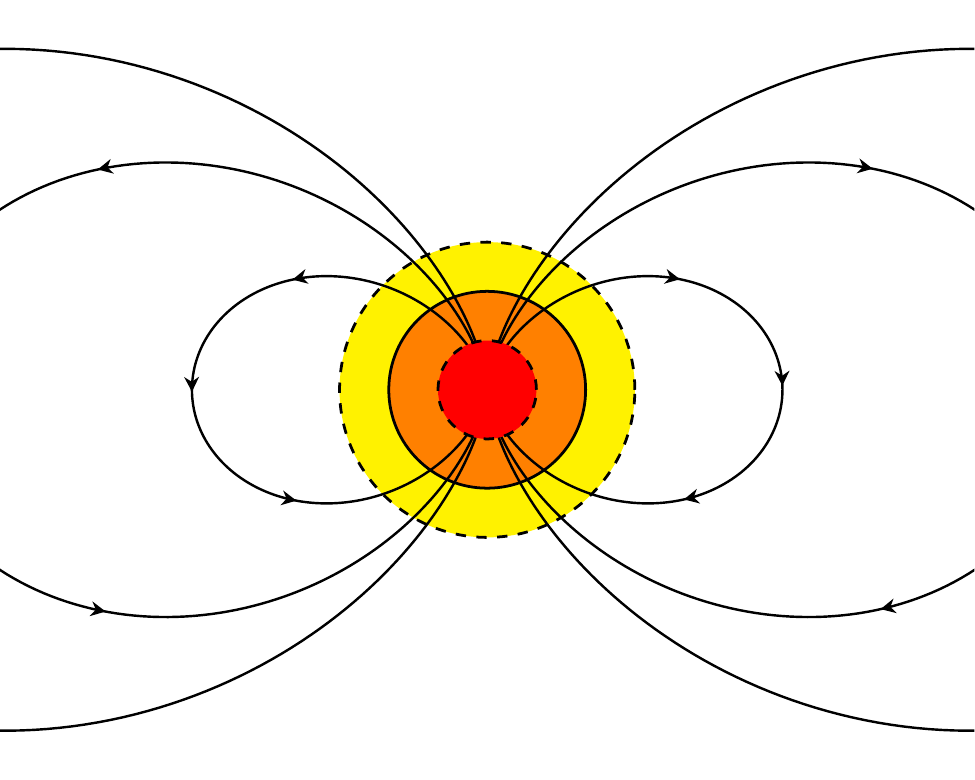}
\caption{Diagram illustrating the three layers used to initialise and hold constant the star or planet. Red indicates the central region of the body, within which all fluid quantities are held constant with time at their initial values with the magnetic field $16 B_{\mathrm{eq}} \hat{z}$. The orange shell is identical to the red region except that the magnetic field has a dipole configuration. Yellow indicates the outer atmosphere of the body in which all quantities except the magnetic field are held constant with time. The solid black circle indicates the surface of the body. 
\label{fig:interior}}
\end{figure}

The internal magnetic field is specified as a series of three concentric shells. The inner region where $0 \ R~<~r~<~0.5 \ R$ has a constant magnetic field of $16 B_{\mathrm{eq}} \hat{z}$. This is done so that the field magnitude in this region meets the magnitude of the dipole field in the second layer, $0.5 \ R~<~r~< 1 \ R$, in a smooth manner. Beyond this radius the field is still dipolar but time dependent and allowed to be modified by the outflowing wind. This configuration avoids the singularity at $r~=~0$ in equation (\ref{eq:b_field}). In the region $1 \ R~<~r~<~1.5 \ R$, the density, pressure and velocity are set and held constant in time according to equations (\ref{eq:vel_start} - \ref{eq:vel_end}) and equations (\ref{eq:pres_init} and \ref{eq:rho_init}) in Section \ref{sec:model_init}, to reflect the initial wind region. The magnetic field is free to evolve in this region and relax into the flow. All three regions which both the star and planet are divided into are illustrated in Fig. \ref{fig:interior}.

\subsubsection{Stellar and Planetary Surface Parameters}
\label{sec:params}

Throughout this study, stellar parameters such as radius, mass and coronal temperature are set to solar values and are summarised in Table \ref{tab:params}. Parameters for the planet are the same as those used by \cite{Matsakos2015} who parametrised the values for base density and pressure using high resolution 1D simulations of the stellar and planetary outflows originally conducted by \cite{Matt2008}. The base values are adapted such that mass-loss rates ($\dot{M}$) and wind values in the 3D simulations match those of the 1D models.

\begin{figure}
\centering
\includegraphics[width=0.5\textwidth]{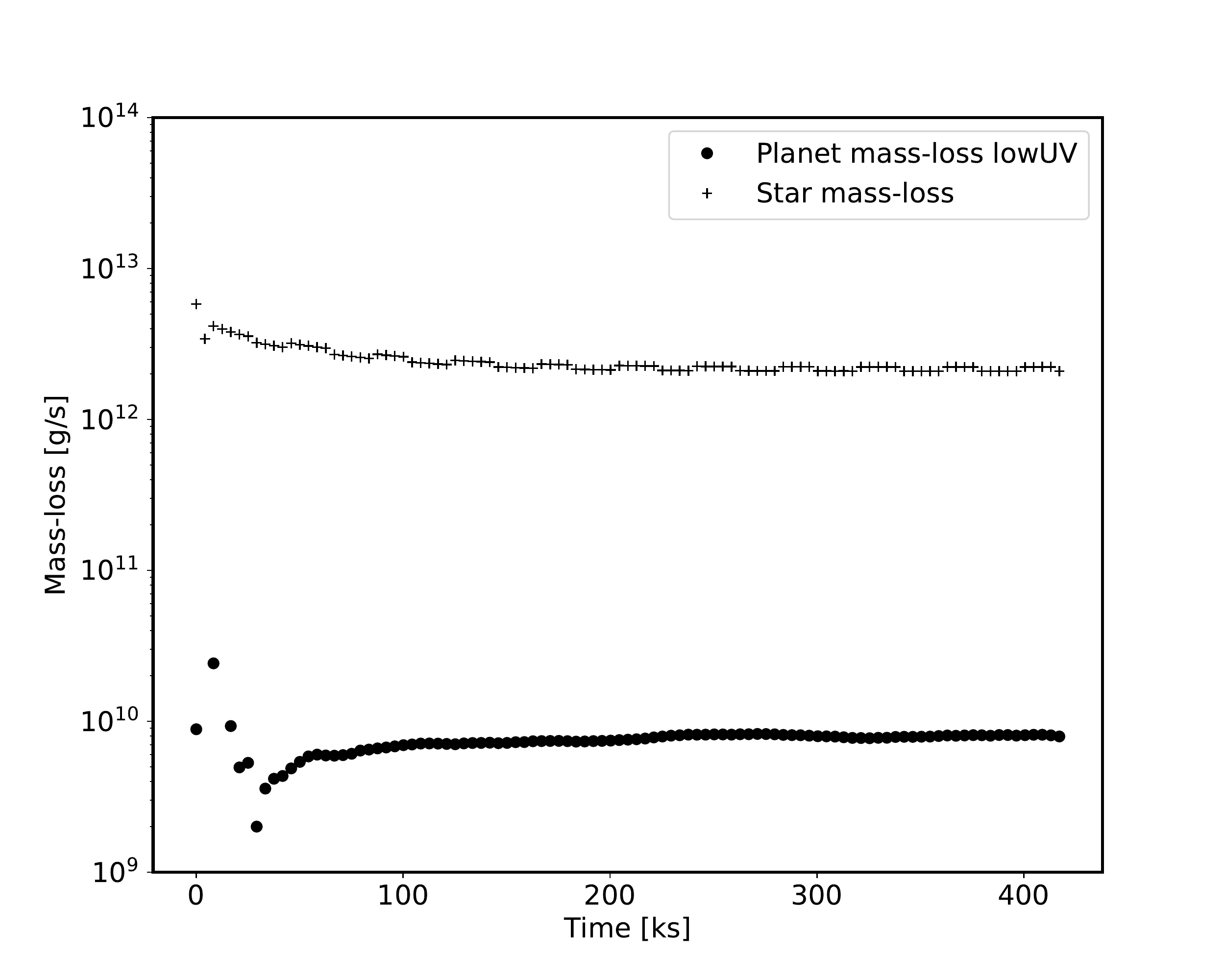}
\caption{Mass-loss rates for the bodies in the simulation. The star has a steady state mass-loss of $2.14 \times 10^{12} \mathrm{g/s}$ and the planet has a steady state mass-loss of $7.9 \times 10^{9} \mathrm{g/s}$. The simulation was run for a total time of $417.2 \mathrm{ks}$ to reach steady state.
 \label{fig:mass-loss}}
\end{figure}

In this study, to verify the above approach, $\dot{M}$ from both the star and planet were measured directly from the simulation results. This was done by specifying a surface of constant radius around the star and planet of $2 R_{\ast}$ and $2 R_{\circ}$ respectively and calculating the flux of material across this surface. The mass-loss rates are shown in Fig. \ref{fig:mass-loss}. The values at the start of the simulation are due to the winds being initialised as non-magnetised outflows. As the winds respond to the presence of the stellar and planetary magnetospheres, $\dot{M}$ readjust as the radial velocity decreases according to mass conservation: $\dot{M}~=~4 \pi \rho r^{2} v(r)$. In the stellar case, $\dot{M}$ reaches a stable value at $2.14 \times 10^{12} \ \mathrm{g/s}$. This values is consistent with the solar $\dot{M}$. The planetary mass-loss reaches equilibrium at $\sim 8 \times 10^{9} \ \mathrm{g/s}$, which is constant with simulations by \cite{Salz2016} for HJs such as WASP-12 b \citep{Hebb2009} and GJ 3470 b \citep{Bonfils2012}. Both of these exoplanets have semi-major axes which are smaller to the value used in this work but have comparable surface temperatures. 

The topic of asymmetric mass-loss between a planets day- and night-side is an active area of research. \cite{Tripathi2015} study this difference and determine a steady state mass-loss of $\sim 2 \times 10^{10} \ \mathrm{g/s}$ for a HJ with similar parameters to those used in this study. This mass-loss is comaprable, within an order of magnitude, to that obtained in the simulation presented here, which assumes spherical mass-loss according to a Parker wind solution. \cite{Tripathi2015} employ a sophisticated treatment of the ionisation balance and the dynamics of neutral plasma species and do not include magnetic fields and focus solely on the region within the direct vicinity of the HJ. For these reasons and because we wish to investigate the global evolution of the evaporated material throughout the star-planet system, we chose to adopt the more simplistic approach described at the start of Section \ref{sec:modelling}.

As both the star and the planet serve as sources of magnetised flow in the simulation, a correct $\dot{M}$ ensures the correct amount of material is replenished in the computational domain, in order to keep the simulation in quasi steady-state, once the initial conditions have advected to the outer boundaries. This does not however guarantee that flow dynamics in the vicinity of the star or planet are correctly reproduced in 3D simulations. Magnetic field geometry plays a large, if not dominant role and will be discussed in Section \ref{sec:Circum}.

The choice of equatorial magnetic field strength, $B_{\mathrm{eq}}$, for the star was based on approximate values for the solar magnetic field, $2 G$ (see Table \ref{tab:params}). In the case of the HJ, we assume that the field is driven by dynamo action due to the rotation of the planet \citep{Stevenson2003, Strugarek2014a}. Since the HJ is tidally locked \citep{Griesmeier2004}, its rotation is much slower than that of Jupiter, which has an $B_{\mathrm{eq}}~\approx~15 \ \mathrm{G}$. As such an $B_{\mathrm{eq}}~=~1 \ \mathrm{G}$ was chosen. This value is consistent with those found in the literature \citep{Pillitteri2015, Strugarek2015, Nichols2016}.

For simplicity, the magnetic field topology of both the star and planet was set to dipolar, with the dipole moment aligned with the rotation axis, which in turn is perpendicular to the plain of the ecliptic. This configuration produces a planetary magnetosphere which corresponds to the anti-aligned case of \cite{Strugarek2015} (anti-aligned with respect to the \emph{local} magnetic field of the stellar wind at the planets position). 

\begin{table*}
\centering
\caption{Stellar and planetary parameters used in the simulations.}
\begin{tabular}{ccccc}
\hline
Parameter & Symbol & Star & Planet & \\
\hline
Mass & $M_{\ast, \circ}$ & $1 \ \mathrm{M_{\sun}}$ & $0.5 \ \mathrm{M_{J}}$ \\
Radius & $R_{\ast, \circ}$ & $1 \ \mathrm{R_{\sun}}$ & $1.5 \ \mathrm{R_{J}}$ \\
Temperature & $T_{\ast, \circ}$ & $1 \times 10^{6} \ \mathrm{K}$ & $6 \times 10^{3} \ \mathrm{K}$ \\
Equatorial magnetic field strength & $B_{eq \ast, \circ}$ & $2 \ \mathrm{G}$ & $1\ \mathrm{G}$ \\
Surface density & $\rho_{\ast, \circ}$ & $5 \times 10^{-15} \ \mathrm{g/cm}^{3}$ & $7 \times 10^{-17} \ \mathrm{g/cm}^{3}$  \\
Orbital radius & $a$ & $-$ & $0.047 \ \mathrm{au}$ \\
Orbital period & $p_{\mathrm{orb}}$ & $-$ & $3.7 \ \mathrm{days}$ \\
Rotational period & $p_{\mathrm{rot} \ast, \circ}$ & $3.7 \ \mathrm{days}$ & $3.7 \ \mathrm{days}$ & \\
\hline
\end{tabular}
\label{tab:params}
\end{table*}

\subsection{Electron Cyclotron Maser Instability}

\subsubsection{Governing Equations}
\label{sec:ECMI}

The history and applications of this theory of emission of electromagnetic radiation via the ECMI process is summarised in a recent review by \cite{Treumann2006}. The frequency at which emission due to the ECMI process propagates at is given by the frequency of gyration of electrons about magnetic field lines; known as the cyclotron frequency:
\begin{equation}
\label{eq:ce}
\nu_{\mathrm{ce}} (\mathrm{MHz}) = \left( \frac{e B}{2 \pi m_{\mathrm{e}} c} 
\right) = 2.80 B, 
\end{equation} 
\citep{Stevens2005}. As described by \cite{Treumann2006}, the ECMI is a plasma instability which, given a background non-thermal electron population, pumps directly the free-space electromagnetic modes. These modes are the result of a dispersion relation for the propagating electromagnetic radiation and leads to several conditions on efficient generation. The relevant modes for the present study is the RX-mode and LO-mode; whose lower cut-off references are
\begin{equation}
\label{eq:cut-off}
\nu_{\mathrm{X}} = \frac{1}{2} \left[ \nu_{\mathrm{ce}} + \left( \nu_{\mathrm{ce}}^{2} + 4 \nu_{\mathrm{pe}}^{2} \right)^{1/2} \right]
\end{equation}
and
\begin{equation}
\label{eq:pe}
\nu_{\mathrm{pe}} (\mathrm{MHz}) = \left( \frac{n_{\mathrm{e}} e^{2}}{\pi 
	m_{\mathrm{e}}} \right)^{1/2} = 8.98 \times 10^{-3} n_{\mathrm{e}}^{1/2}.
\end{equation}
respectively. Finally, electrons contributing to the ECMI process must also follow a loss-cone distribution function, which is the case for magnetospheric cusp configurations such as the one discussed here.

If the two frequencies, equations (\ref{eq:cut-off} and \ref{eq:pe}) are exceeded and
\begin{equation}
\label{eq:ECMI}
\frac{\nu_{\mathrm{ce}}}{\nu_{\mathrm{pe}}} > 1,
\end{equation}
then there is non-negligible emission due to the ECMI process. In practice, the equation (\ref{eq:ECMI}) needs to be $\gtrsim 2.5$ for the emission to be efficient \citep{Weber2017}. Equation (\ref{eq:ECMI}) will form the diagnostic for measuring cyclotron emission from our simulation results.

In this study, radio ECMI emission is assumed to follow a straight line from the source of emission to the observer. A rigours treatment of emission propagation requires the consideration of the changing refractive index of the stellar wind due to density fluctuations and the $1/r^{2}$ dependence of equation (\ref{eq:rho_init}) along the line emission, modifying it's path, as it travels through the stellar wind. However, this influence on the emission trajectory is assumed to be small and the emission takes a straight line from it's origin to the observer.

\begin{figure}
\centering
\includegraphics[width=0.4\textwidth]{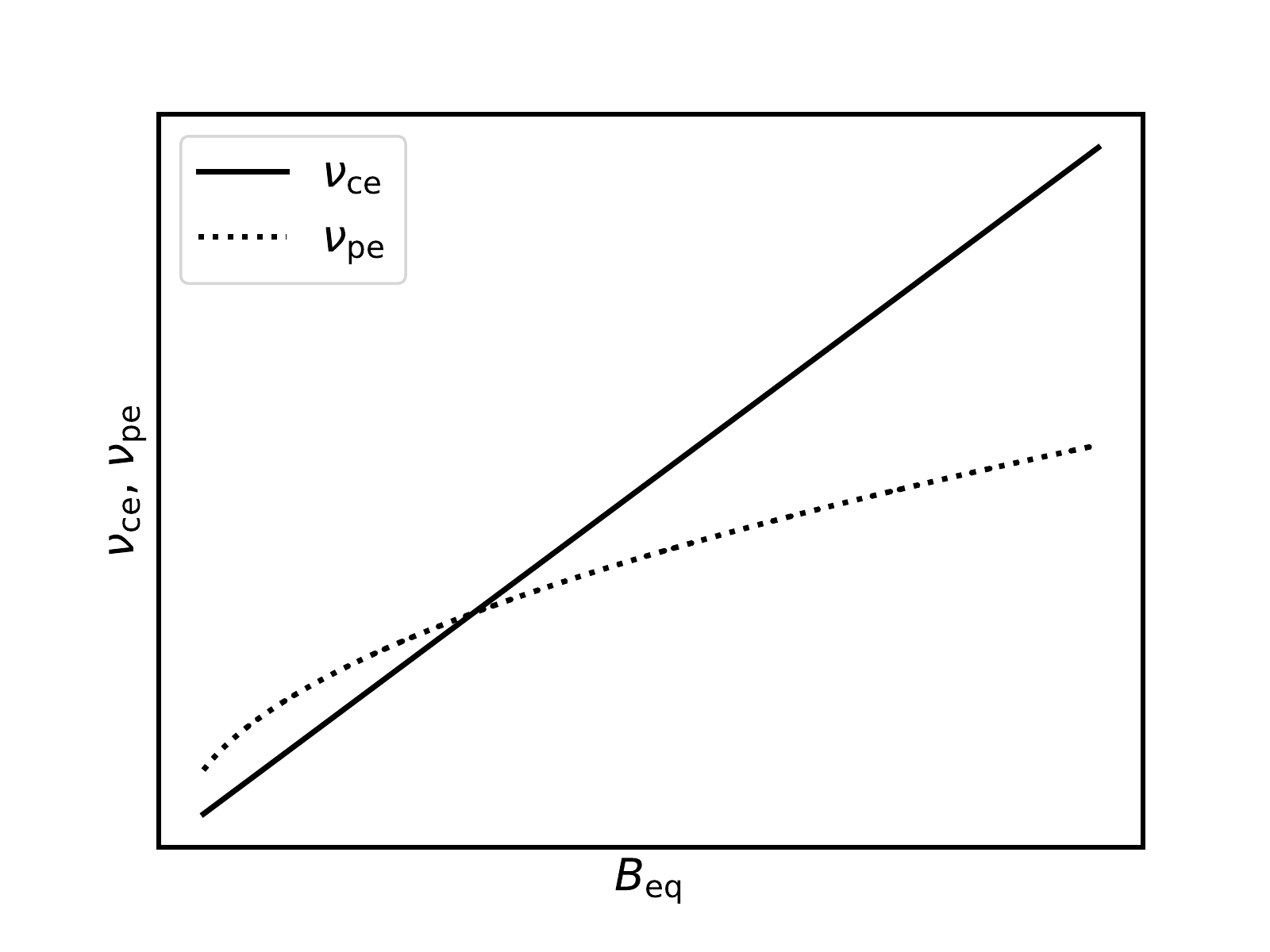}
\caption{Schematic diagram of the scaling for both $\nu_{ce}$ and $\nu_{pe}$ with $B_{\mathrm{eq}}$ at an arbitrary point in space. Assuming that $n_{\mathrm{i}}$ increases linearly with $B_{\mathrm{eq}}$ then the relation $\nu_{\mathrm{pe}}~\propto~\sqrt{n_{\mathrm{i}}}$ means that, at a given magnetic field strength, $\nu_{ce}$ will exceed $\nu_{pe}$ and ECMI will be efficient at producing radio emission.
\label{fig:frequency_scalling}}
\end{figure}

Equation (\ref{eq:ce}) is a linear function of the local magnetic field strength, $B$. Therefore, if the planetary $B_{\mathrm{eq}}$ were to increase, so to would $\nu_{\mathrm{ce}}$. For a fixed mass-loss rate, increasing $B_{\mathrm{eq}}$ would result in greater confinement of material in the immediate vicinity of the planet and an increase in $n_{\mathrm{i}}$. This is due to the greater number of closed field lines and larger volume of the planetary magnetosphere, trapping a greater number of particles. As equation (\ref{eq:pe}) is a week function of $n_{\mathrm{i}}$, $\nu_{pe}$ would also see an increase but to a lesser degree than $\nu_{\mathrm{ce}}$. Therefore, $\nu_{\mathrm{ce}}$ and $\nu_{\mathrm{pe}}$ both scale with $B_{\mathrm{eq}}$. For example, if $B_{\mathrm{eq}}~\rightarrow~2 B_{\mathrm{eq}}$ then $\nu_{\mathrm{ce}}~\rightarrow~2 \nu_{\mathrm{ce}}$ while $\nu_{\mathrm{pe}}~\rightarrow~\sqrt{2} \nu_{\mathrm{pe}}$. but only if doubling $B_{\mathrm{eq}}$ resulted in twice the number of confined ions. This simplified theoretical scaling of $\nu_{\mathrm{ce}}$ and $\nu_{pe}$ with $B_{\mathrm{eq}}$ is shown in Fig. \ref{fig:frequency_scalling}. Making the assumption that doubling $B_{\mathrm{eq}}$ leads to twice the number of confined ions, therefore doubling $n_{\mathrm{i}}$, we can see that as $B_{\mathrm{eq}}$ increases, both $\nu_{\mathrm{ce}}$ and $\nu_{pe}$ (via $\nu_{\mathrm{pe}}~\propto~\sqrt{n_{\mathrm{i}}}$) do also and there is a value of $B_{\mathrm{eq}}$ where equation (\ref{eq:ECMI}) exceeds unity and ECMI emission is produced.

\subsubsection{Emission Generation}
\label{sec:emission}

\begin{figure}
\centering
\includegraphics[width=0.49\textwidth, trim={0 0 0 0},clip]{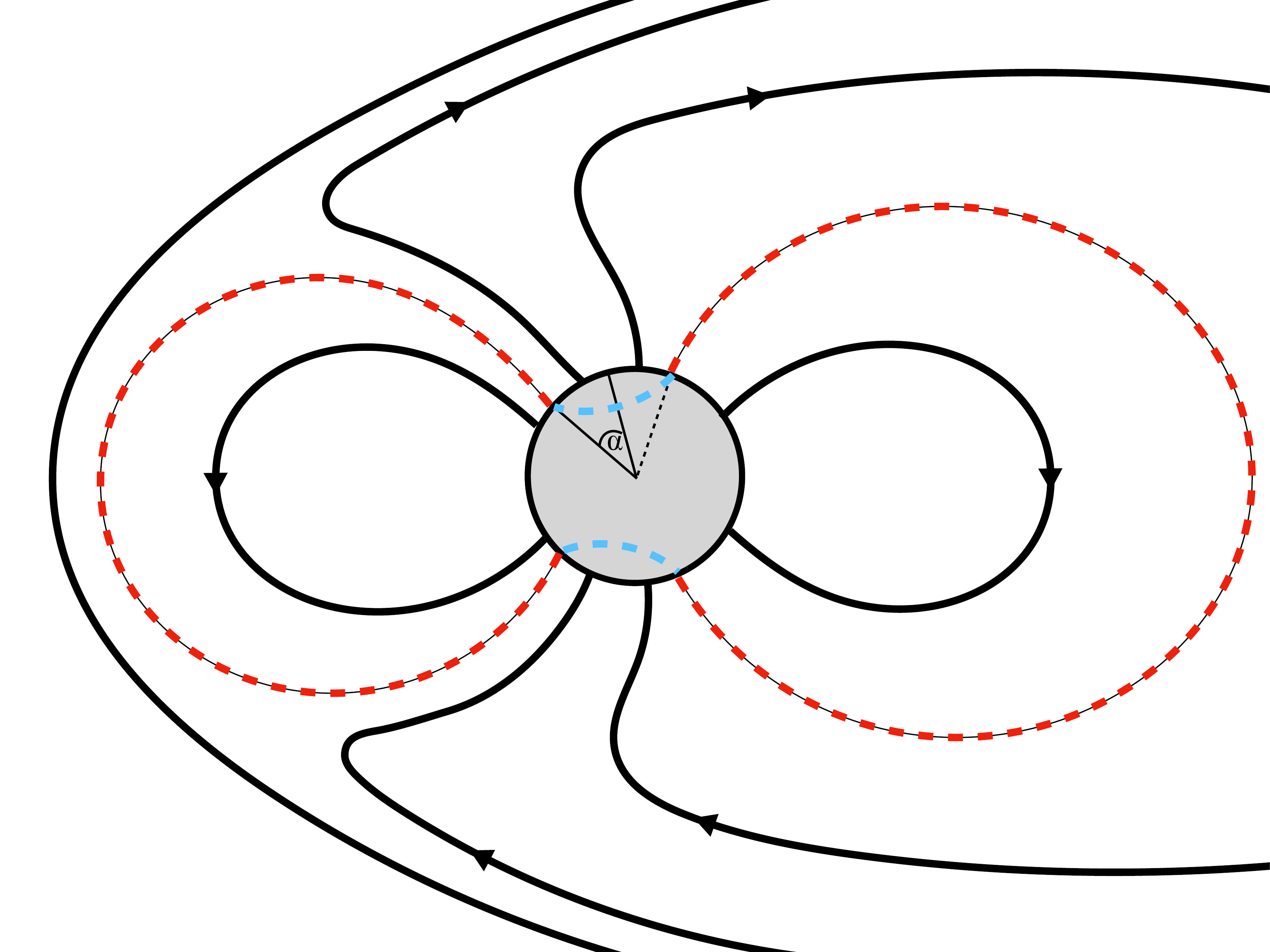}
\caption{Schematic diagram of the magnetic field geometry for a HJ embedded in a stellar wind. To the left, the solid black line with no arrows indicates the bow shock. The longest closed field lines are highlighted in red. Field lines at higher latitudes are effectively open to the stellar wind. The angle $\alpha$ indicates the lowest latitude above which the field lines are open. The dashed blue lines are the regions from which the ECMI emission is expected.
\label{fig:field_lines_latitude}}
\end{figure}

ECMI emission is highly directional. Fig. \ref{fig:field_lines_latitude} shows a diagram of the magnetosphere of an exoplanet embedded in a stellar wind. The image illustrates active regions where magnetic field lines form either a closed or open magnetosphere. The red highlighting show the longest closed field lines which form exoplanetary analogies with the Van Allan radiation belts in the Earth's magnetosphere. Magnetic field lines between these two closed regions form the open part of the magnetosphere allowing electrons from the stellar wind to enter. Again, in analogy with Earth, these regions are known as the polar cusps. Electrons entering the cusps process in a helical path whose frequency of gyration is directly proportional to the magnetic field strength (see equation (\ref{eq:ce})). Any emission produced by this motion is directed into a hollow conical opening pointing in the direction of electron motion. This results in highly directional emission \citep{Dulk1985}. \cite{Vidotto2011} calculated the $\nu_{\mathrm{ce}}$ using the opening angle, $\alpha$,  corresponding to the latitude of the closed magnetic field lines marked in red in Fig. \ref{fig:field_lines_latitude}, giving an estimate of the shape of the emitting region. The field strength at this latitude gives $\nu_{\mathrm{ce}}$ and $\alpha$ can be approximated by:
\begin{equation}
  \alpha = \arcsin{\left[ \left( R_{\circ} / R_{\mathrm{M}} \right)^{1/2} \right]}.
  \label{eq:alpha}
\end{equation}
$R_{\mathrm{M}}$ is the radius corresponding to the largest closed field line and is inferred from simulation results in Section \ref{sec:radio_p}. This together with equation (\ref{eq:ce}) gives $\nu_{\mathrm{ce}}$.

\cite{Tilley2016} calculate the directional dependence of the ECMI flux emitted from the poles of a simulated HJ, as projected onto a celestial sphere centred on the exoplanet. directional dependence was found to be highly influenced by the stellar wind ram pressure and planet magnetosphere topology. However, they do not calculate absolute flux levels, only showing normalized fluxes. We do not calculate the direction of the emission here as this study is principally concerned with the plasma environment in the stellar and planetary wind and whether and where the ECMI emission is produced in the first place.

While $\nu_{ce}$ is dependent on the magnetic field strength and geometry, the intensity of emission is a function of the power available to the electrons traped in the magnetosphere. This power is estimated by \cite{Zarka1998} (and later used by \cite{Stevens2005}) as
\begin{equation}
  P_{\mathrm{r}} = \delta \frac{ \dot{M}_{\ast} u_{\mathrm{W}}^{2} R_{\mathrm{eff}}^{2}}{ 4 a^{2}}.
  \label{eq:radio_p}
\end{equation}
Here, $\delta$ is an efficiency parameter ($\sim 7 \times 10^{-6}$), $\dot{M}_{\ast}$ the stellar mass-loss rate, $u_{\mathrm{W}}$ the stellar wind at the orbital radius, $a$ and $R_{\mathrm{eff}}$ is the effective radius of the magnetosphere as seen by the stellar wind and the only factor in equation (\ref{eq:radio_p}) that is a property of the exoplanet. The star plays the central role in supplying power in the form of incident kinetic and magnetic energy. For a typical HJ $P_{\mathrm{r}} \sim 10^{15} \ \mathrm{W}$ \citep{Stevens2005, Zarka2007}

Equation (\ref{eq:radio_p}) assumes that the result of equation (\ref{eq:ECMI}) is greater than the threshold for efficient production of radio emission. In Section \ref{sec:Res} it will be shown that this simple model is insufficient to accurately describe the ECMI process in HJs.

\subsection{Numerical Modelling}
\label{sec:Num}

The computational mesh is initialized everywhere according to the equations presented in Section \ref{sec:models} for the stellar wind. A region of $10 R_{\circ}$ around the planet is initialized for the planetary wind in the same manner. The MHD equations (equations (\ref{eq:mass} - \ref{eq:magnetic})) were solved numerically using the publicly available MHD code PLUTO (version 4.2) \citep{Mignone2007, Mignone2011}. This was done using a 2nd order scheme with linear spatial reconstruction (Van Lear limiter) and 2nd order Runga-Kutta for the time stepping. This was paired with the HLLD Riemann solver. The magnetic field zero divergence condition was enforced using the GLM formalism of \cite{Dedner2002}, see also \cite{Mignone2010, Mignone2010a} for it's implementation in PLUTO.

The outer boundary conditions to the computational domain are set to zero gradient (outflow). As the velocity every where just inside the boundary is directed outwards, these boundary conditions insure that the finite computational domain, does not influence the solution.

The vicinity of the planet is characterised by a high magnetic field strength and a high density but low velocity outflow. This results in a plasma-$\beta$ $<<~1$. Therefore the magnetic field dominates the evolution of the flow in this region. These conditions can prove challenging to the numerical scheme and lead to the development of unphysical structure within the planetary magnetosphere. This issue is alleviated by using the technique of background field splitting in which equations \ref{eq:mass} - \ref{eq:magnetic}) are rewritten with the magnetic field $\bm{B}_{\mathrm{tot}}~\rightarrow~\bm{B}_{0} + \bm{B}_{1}$. Where the total magnetic field, $\bm{B}_{\mathrm{tot}}$, is split into an initial field, $\bm{B}_{0}$, which is static with respect to time, and a perturbation to this initial field, $\bm{B}_{1}$, which is evolved with the simulation. This allows for the planetary outflow to smoothly interact with the magnetosphere as it expands from the planet surface, reducing grid aligned flow artefacts. Using this technique results in a stable time step, remaining constant throughout the simulation.

To ensure that the simulation reaches quasi-steady state, the simulation is evolved through $\times~10$ the time for a fluid particle to be advected from the surface of the star to the outer boundary. This flow time is approximately 10 hours, therefore, the simulation is evolved through $100~\mathrm{hours}$ or $360 \ \mathrm{ks}$.

The following sections detail the specifics of the numerical mesh used in the simulation and the dynamically adaptive refined mesh used. In the following, $x$, $y$ and $z$ are relative to the origin of the coordinate system.

\subsubsection{Simulation Mesh}

The Physical extent of the mesh employed in our simulation was $-32 \ R_{\ast}~<~x, y~<~32 \ R_{\ast}$ and $-16 \ R_{\ast}~<~z~<~16 \ R_{\ast}$. This region was discretised into a mesh with an initial resolution of $128^2$ cells in the $xy$ plane and 64 cells in the $z$ direction. This gives a resolution of 2 cells per stellar radius. This initial mesh was successively refined with a maximum 5 AMR levels to an effective resolution of $4096^2 \times 2048$ or 64 cells per stellar radius.

\begin{figure}
\centering
\includegraphics[width=0.48\textwidth]{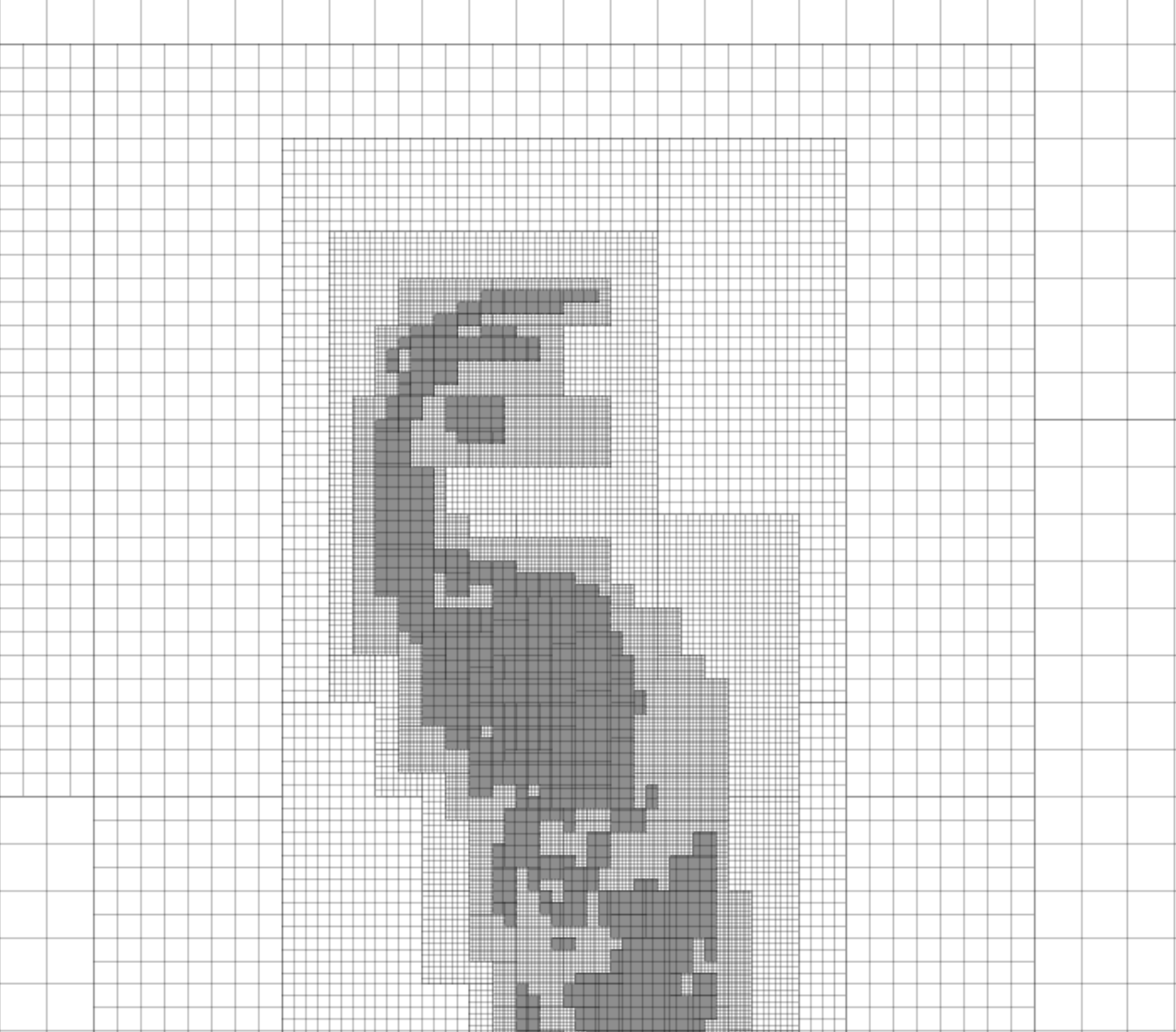}
\caption{Example of the mesh refinement employed in the simulation. The patch regions denote different refinement levels. Each additional refinement level represents a two times increase in the resolution from the previous. The most coarse level can be seen on the left and right of the image and is the base resolution and the highest refined region, centred on the planet and bow shock.
 \label{fig:Mesh}}
\end{figure}

The mesh refinement is carried out by checking the gradient of the density in each cell at every second time step. If the density gradient fulfils a threshold criteria then the cell is marked for refinement, see \cite{Mignone2011} for details. This allows the mesh to track the evolution of the planetary material as it interacts with the stellar wind. An example of the refined mesh is shown in Fig. \ref{fig:Mesh}. Each nested patch indicates an additional AMR level which is twice as refined as it's parent patch.

\subsubsection{Assumptions and Limitations}

The base of the stellar and planetary winds are held constant out to $1.5 \ R_{\ast}$ and $1.5 \ R_{\circ}$ respectively. This is done to allow the stellar and planetary outflows to initialise properly and not stall in the presence of the magnetic field of either body. This introduces the limitation that any material, either form the star or planet, can only come within 1.5 $R_{\ast}$($R_{\circ}$) of the star(planet) surface. This limits the ability to use the simulation results presented here for studying accretion of material onto either body. It is left to future studies to relax this condition and to study accretion in greater detail.

\begin{figure}
\centering
\includegraphics[width=0.49\textwidth]{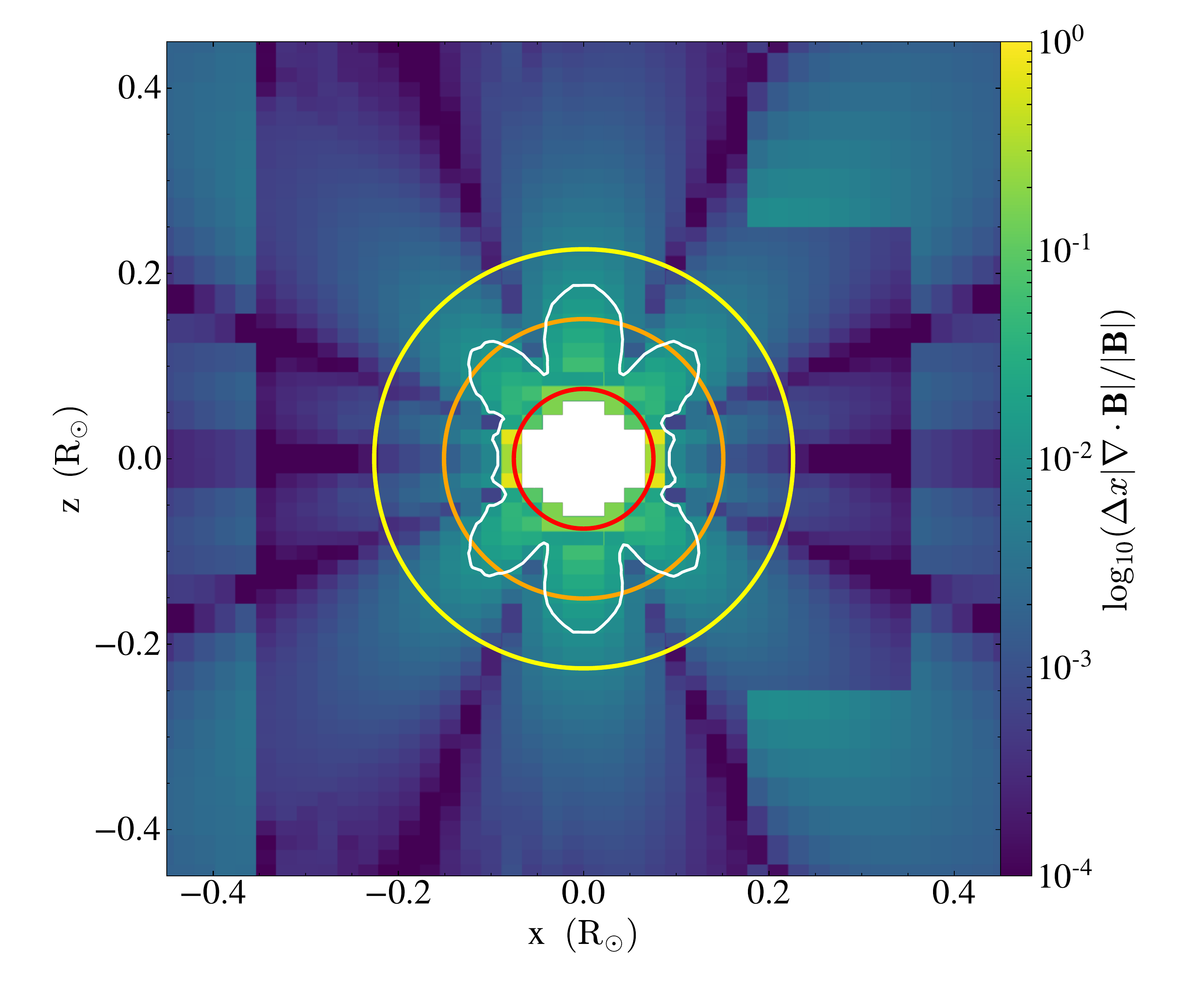}
 \caption{Plot of the magnetic field divergence measure $\log_{10} \left( \Delta x | \nabla \cdot \bm{B} | / | \bm{B} | \right)$. The three circles and their colouring indicate the layers of the planet as described in Section \ref{sec:Interiors}. A single white contour indicates the surface within which $\log_{10} \left( \Delta x | \nabla \cdot \bm{B} | / | \bm{B} | \right) > 10^{-2}$. At all points, this region is contained within the region where the solution is held constant with time, as such the divination from $\nabla \cdot \bm{B} = 0$ has no influence here. Outside the planet $\log_{10} \left( \Delta x | \nabla \cdot \bm{B} | / | \bm{B} | \right) < 10^{-2}$, a value considered to have minimal effect of the solution.} 
\label{fig:divB}
\end{figure}

A central consideration in numerical MHD simulation is insuring the $\nabla \cdot \bm{B} = 0$ condition. The configuration in Section \ref{sec:Interiors} for the internal magnetic field presents a challenge. As the solution is over written at each time step, mono-poles are introduced at the interface between the constant field (internal) and the evolving field (external). The extent to which this influences the evolution of the simulation can to some extent be messured using the approach of \cite{Hopkins2016}, where a measure of the divergence is given by $\Delta x | \nabla \cdot \bm{B} | / | \bm{B} |$, with $\Delta x$ the cell width. To assess the impact of the static planetary core (and by extension the static stellar core), this quantity is plotted in Fig. \ref{fig:divB} for a slice plot of the steady state solution, aligned with the $z$-direction, intersecting the centre of the planet. $\Delta x | \nabla \cdot \bm{B} | / | \bm{B} | < 10^{-2}$ in all regions exterior to the planet. Higher values are contained to the interior, static core of the body, allowing us to draw the conclusion that the divergence cleaning method together with the static planetary interior, does not introduce systematic errors into the solution.

An alternative method is to assign a large enough density value to the planet such that the magnetic field diffusion time scale is much greater than the simulation time scale. Since the planetary wind is defined by the inflow density, fixed on the surface, we are not free to used this method.

One further major assumption is that the entire star and planetary wind material is fully ionised and therefore optically thin to the UV radiation from the star, incident on the planet and providing the source of the atmosphere's photoevaporation. This assumption could only be relaxed if the planetary outflow was not fixed at its base and the photoevaporation was calculated based on radiation transfer. Such a calculation is beyond the scope of this study. 

Simulation results including initial conditions, global and circumplanetary evolution and ECMI efficiency calculations are presented in the following sections.

\section{Results and Discussion}
\label{sec:Res}

The results are laid out in the following manner: global simulation initial conditions, evolution and steady state solution, description of the magnetised stellar wind, description of the circumplantary environment, quantification of the ECMI radio power and frequency of emission and finally calculation of $\nu_{\mathrm{ce}} / \nu_{\mathrm{pe}}$ with the determination of the efficiency for the generation of radio emission via the ECMI process.

\subsection{Initial Conditions}

\begin{figure}
\centering
\includegraphics[width=0.49\textwidth]{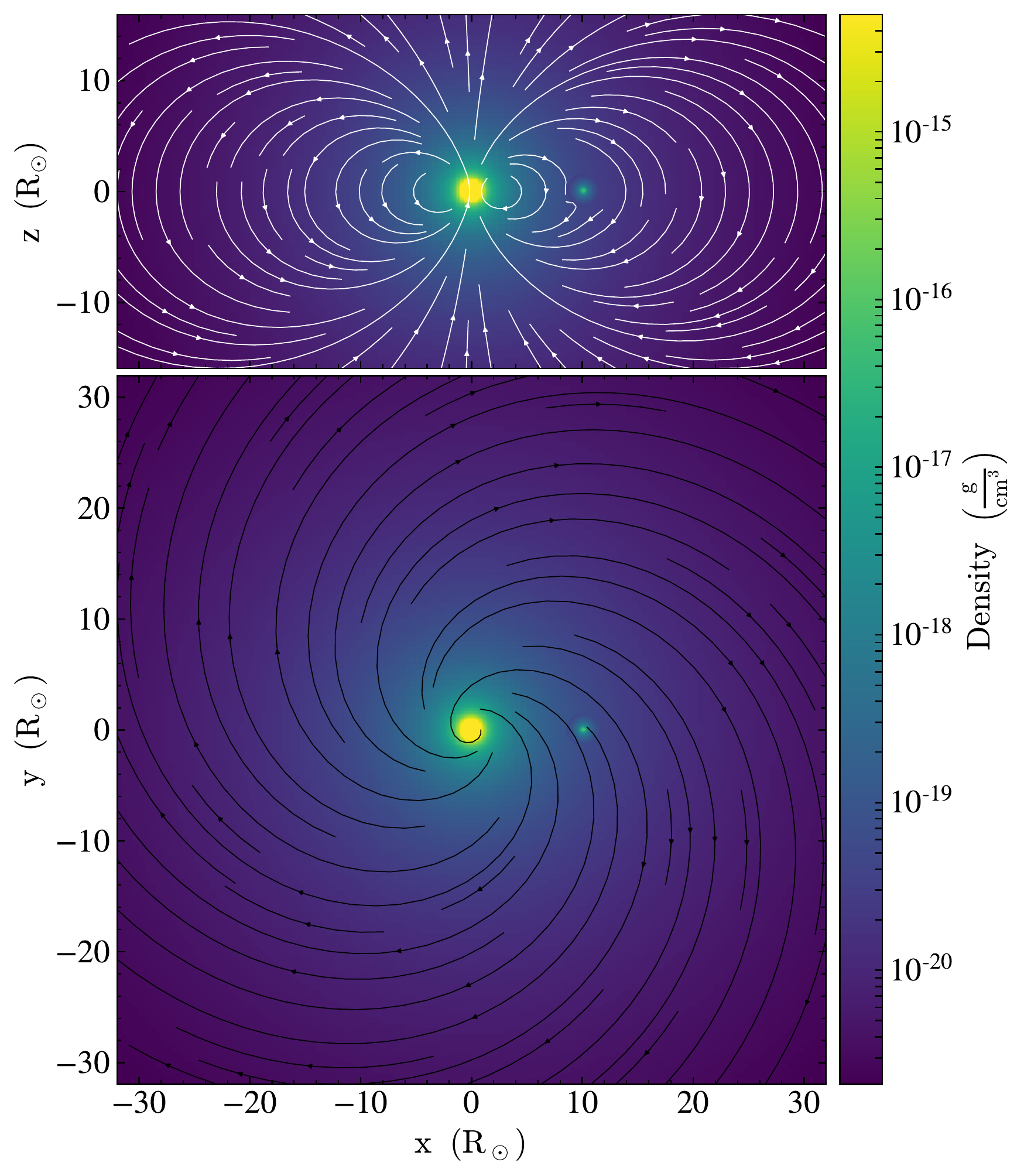}
 \caption{
Initial conditions for the stellar and planetary bodies and the extended wind. Top: view along the y-axis showing the star, planet and total magnetic field which is dominated by the stellar magnetosphere. Bottom: similar to the top image but looking down the z-axis onto the orbital plane, the velocity flow lines of the initial Parker wind model are shown as black lines.
}
\label{fig:initial_conditions}
\end{figure}

The entire simulation domain, star at the centre and planet situated to the right, is shown in Fig. \ref{fig:initial_conditions}. Across the entire simulation domain, the wind initial conditions are set according to the stellar properties only, except for a cavity of radius $10 \ R_{\circ}$ around the planet, which is initialised solely with the planetary wind parameters. This allows the planetary wind to initialise properly and not stall due to the ram pressure of the stellar wind. These initial conditions are designed such that the planetary wind is at approximately terminal velocity and will expand radially in a smooth manner in the first time step, allowing quasi-steady state to be reached in the shortest time possible. As mentioned in Section \ref{sec:Num}, the magnetic field is evolved using the method of background field splitting. As the static background field is required to be force-free, it is initialised as the sum of the stellar and planetary fields, under the assumption that the sum of two force-free fields is itself force-free. 

\subsection{Global Evolution}
\label{sec:global}

\begin{figure}
\centering
\includegraphics[width=0.49\textwidth]{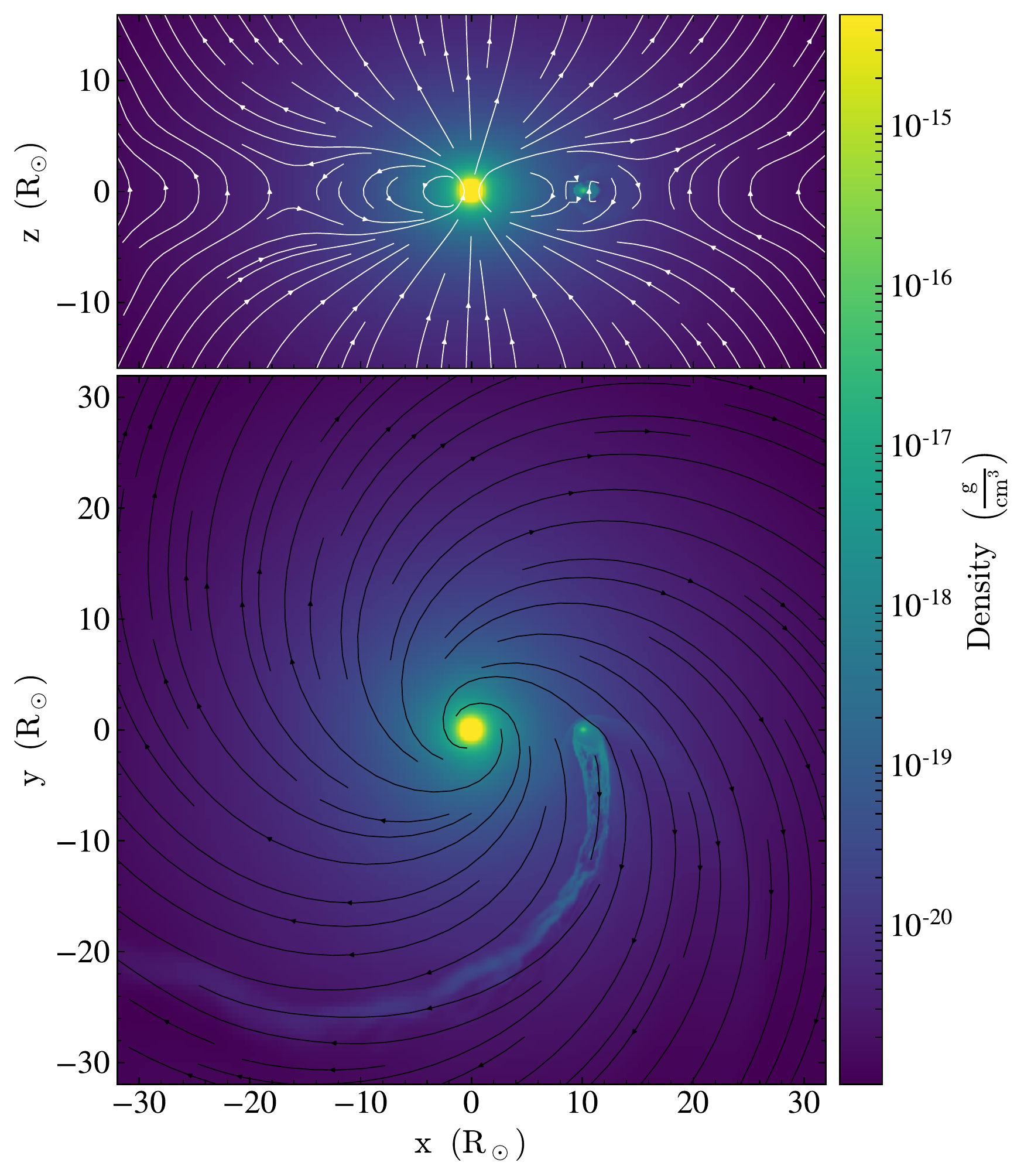}
\caption{
Global quasi-steady state solution, large scale structure of the flow will not change, only short, small scale changes will continue. Top: side on view, looking down the y-axis with overlaid magnetic field lines. Bottom: top down image, looking down the z-axis, of the system with overlaid velocity flow lines. 
\label{fig:steady_state}}
\end{figure}

Fig. \ref{fig:steady_state} shows the final quasi-steady state solution. The top image shows a side on view, looking down the y-axis with overlaid magnetic field lines. The bottom image shows a top down view, looking down the z-axis, of the system with overlaid velocity flow lines. Care must be taken when interpreting the magnetic field lines in the top image, as the field lines shown are constructed using the magnetic vectors in the plane of the slice used to produce the image only. As such, the warping of the field lines by the stellar wind is not fully captured and can lead to artificial structures in the magnetic topology. Therefore, the field lines in in Fig. \ref{fig:steady_state} should be interpreted as illustrative rather than literal. 

At the orbital radius, the stellar wind has drawn open the stellar magnetic field lines to present an open magnetosphere to the planet. The planetary material itself is swept back to form a cometary tail.This material gradually dissipates and is eventually advected into the lower y-boundary of the simulation. The majority of the simulation domain remains dominated by the stellar wind.

Cometary tail structures are also found in the work of \cite{Bourrier2013}, \cite{Bourrier2016} and \cite{Schneiter2016}. The former uses a particle model together with a theoretical Lyman-$\alpha$ absorption line to track the motion of the UV photoevaporated planetary material. They find that for HD209458b, the synthetic and observational absorption profiles agree, leading to a theoretical mass-loss rate between $10^{9} - 10^{11} \ \mathrm{g}/\mathrm{s}$, with the host star producing 3 - 4 times the solar value for ionising flux. This mass-loss rate range corresponds well to the mass-loss determined for the model planet presented here. 

The large temperature difference between the stellar wind and the planetary material between $10^{6} \mathrm{K}$ and $6 \times 10^{3} \mathrm{K}$, suggests that additional physics not incorporated into the simulations presented here could play a substantial role in the evolution of the cometary tail, for example inclusion of thermal conduction or sub-grid turbulence may lead to a higher degree of dissipation. Behaviour of the tail does not play an important role in the ECMI emission, especially in comparison to the planetary polar regions, where the magnetic field is strongest. As such, these additional physics are left to future investigations.

\subsection{Circumstellar Evolution}
\label{sec:Circum}

The magnetic field lines from the stellar surface fall into two categories; either field lines which form closed loops back to the opposite hemisphere or field lines which are open and would in reality connect with the interplanetary/interstellar magnetic field, with a cutoff between these open and closed field lines at a specific line of latitude on the stellar surface.  

\begin{figure}
\centering
\includegraphics[width=0.49\textwidth, trim={0 0 0 0},clip]{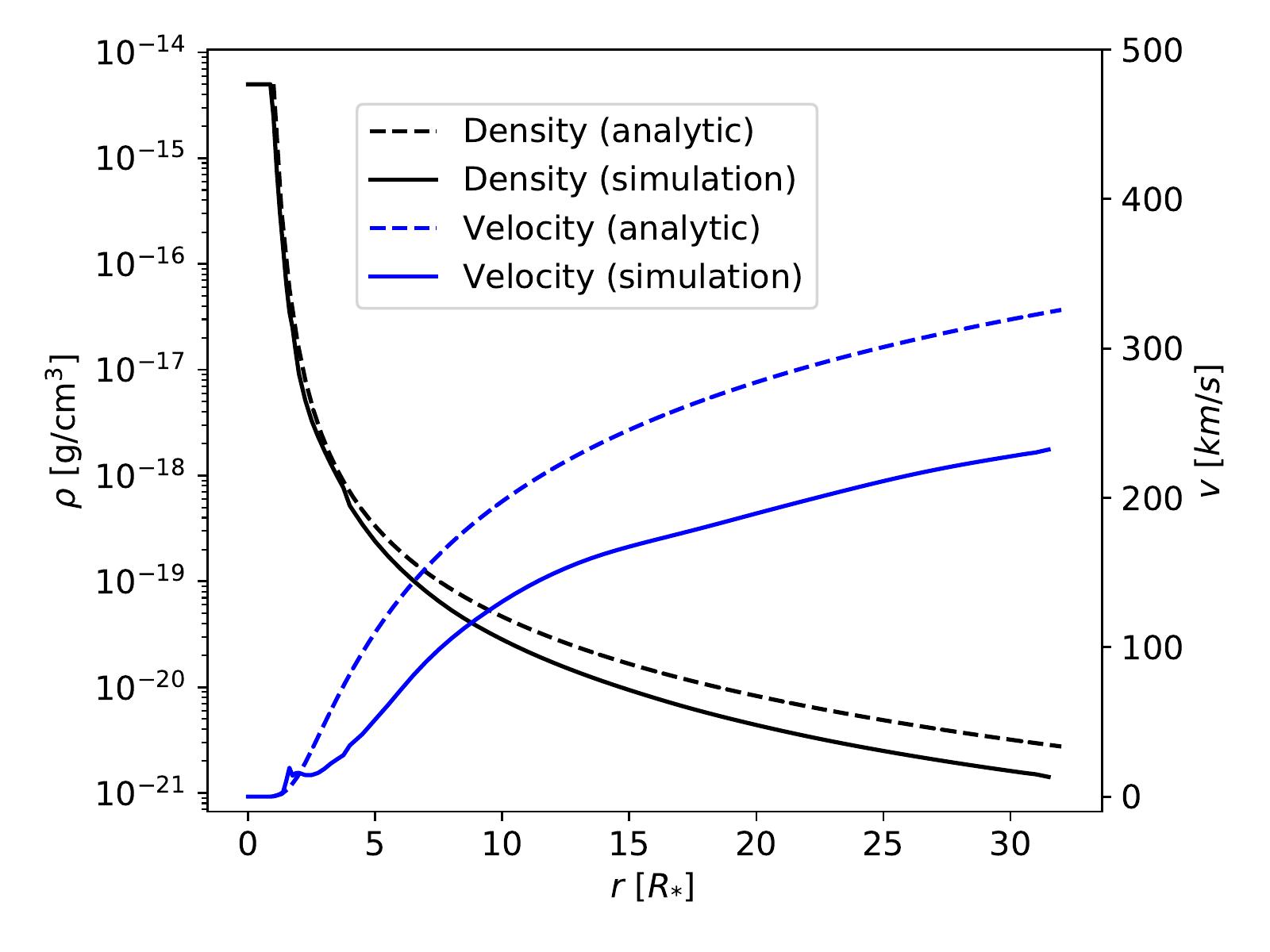}
\caption{Density and velocity of the simulated stellar wind. Shown are the profiles in the orbital plane, characterising the conditions encountered by HJ as it processes on its orbit. Outside of this region the stellar wind will be different due to the dipolar nature of the magnetic field. Dashed lines indicate the analytic solution, as calculated from Parker's equation (equation (\ref{eq:parker_eq})), and solid lines are taken from the steady state simulation results. Both quantities exhibit a decrease with respect to their analytic counterparts, which becomes more pronounced with radial distance form the stellar surface. 
\label{fig:parker_res}}
\end{figure}

To test the validity of the numerical stellar wind results, analytic solutions to Parker's equation (equation (\ref{eq:parker_eq})) along side numerical results of the velocity and density profiles from the simulation are shown in Fig. \ref{fig:parker_res}. Both of the simulated profiles are reduced with respect to the analytic results. This can be seen in the evolution of the stellar mass-loss, discussed in Section \ref{sec:params}, showing initially $\dot{M}_{\mathrm{\ast}} = \dot{M}_{\mathrm{\sun}}$ which then decreases and approaches $2 \times 10^{12} \mathrm{g/cm}^{3}$, a decrease of approximately $30\%$. Closed field lines in the equatorial region of the stellar wind constrain the expanding wind, reducing the velocity and thus the amount of material which can escape the stellar surface. This material then acumulates in the magnetosphere. \cite{Matsakos2015} characterised this accumulation region as the \textit{stellar dead zone} within which the wind can not overcome the combination of stellar gravity and magnetic tension. This region can be seen for $v(r~<~4 R_{\ast})$ in Fig. \ref{fig:parker_res}. This deviation from the Parker solution, agrees with the expected behaviour for a magnetised stellar wind \citep{Matt2008, Matsakos2015}.

\subsection{Circumplanetary Evolution}

To classify the extended atmospheres of HJs \cite{Matsakos2015} divided simulated behaviour into four categories: 
\begin{itemize}
\item type-1; bow shock and thin tail
\item type-2; colliding winds and tail
\item type-3; strong planetary-wind, accretion and tail
\item type-4; Roche-lobe overflow, accretion and tail
\end{itemize}
The behaviour of the HJ atmosphere simulated in this study falls into classification type-1. 

\begin{figure*}
\centering
\includegraphics[width=0.9\textwidth, trim={0 2cm 0 2cm},clip]{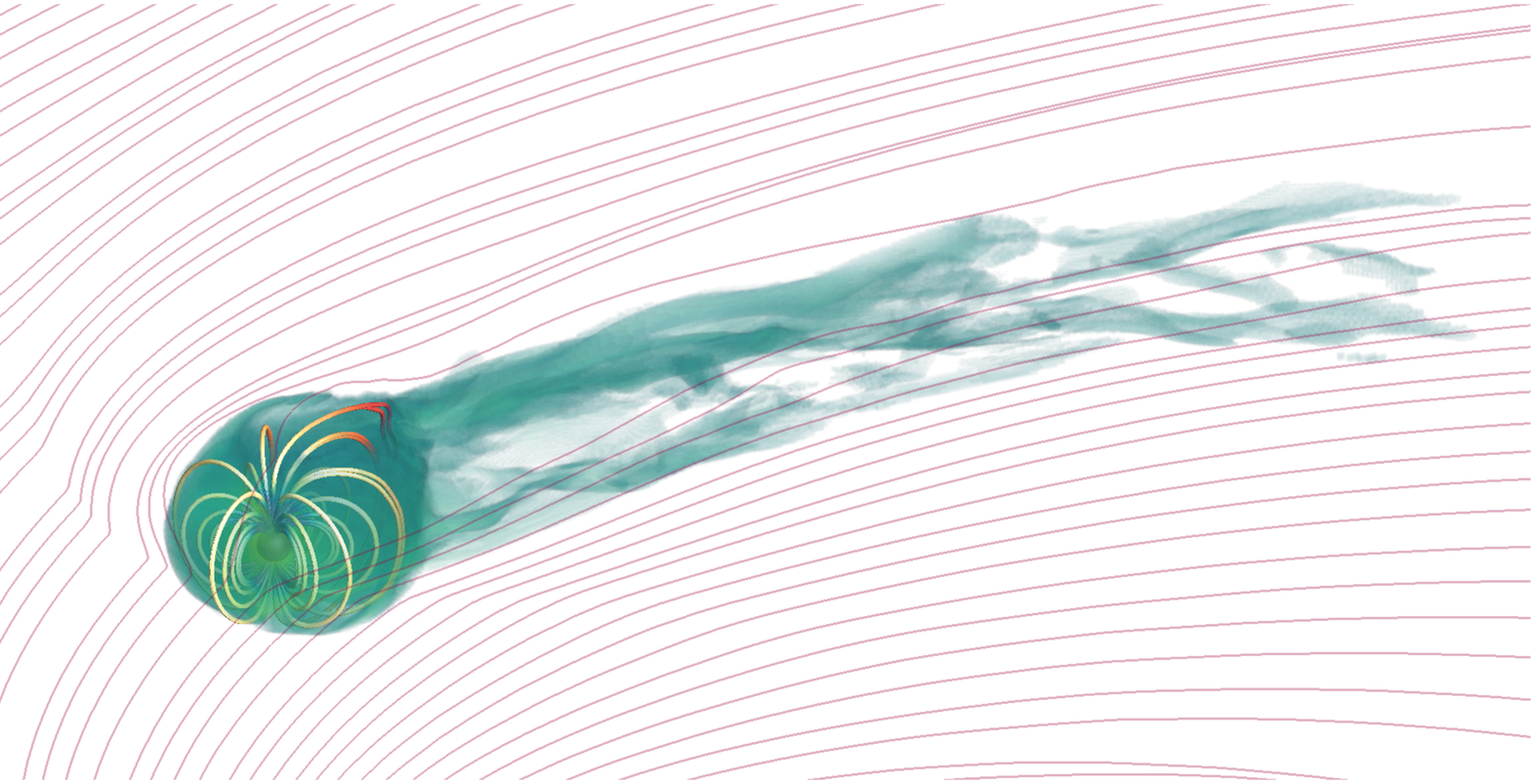}
 \caption{
Volume rendering of the planet. The diffuse green cloud represents density for values greater than $3 \times 10^{-18}$ g/cm$^{3}$.This material forms a cloud that has thermally expanded to fill the planetary magnetosphere and then begun to overflow and mix with the oncoming stellar wind (illustrated here by the red flow lines). The magnetic field lines of the planetary magnetosphere are show with a blue to red colour scheme depicting the field strength.
\label{fig:volume}}
\end{figure*}

A three-dimensional rendering of the planet is shown in Fig. \ref{fig:volume}. The green volume, represents density values greater than $3 \times 10^{-18}$ g/cm$^{3}$. This material surrounds the HJ, fills the magnetosphere by following filed lines (blue to red colour scheme depicting the field strength) and thermally expands until it overflows from the aft part magnetosphere and mixes with the downstream stellar wind.

\begin{figure*}
\centering
\includegraphics[width=0.69\textwidth, trim={0 0 0 0},clip]{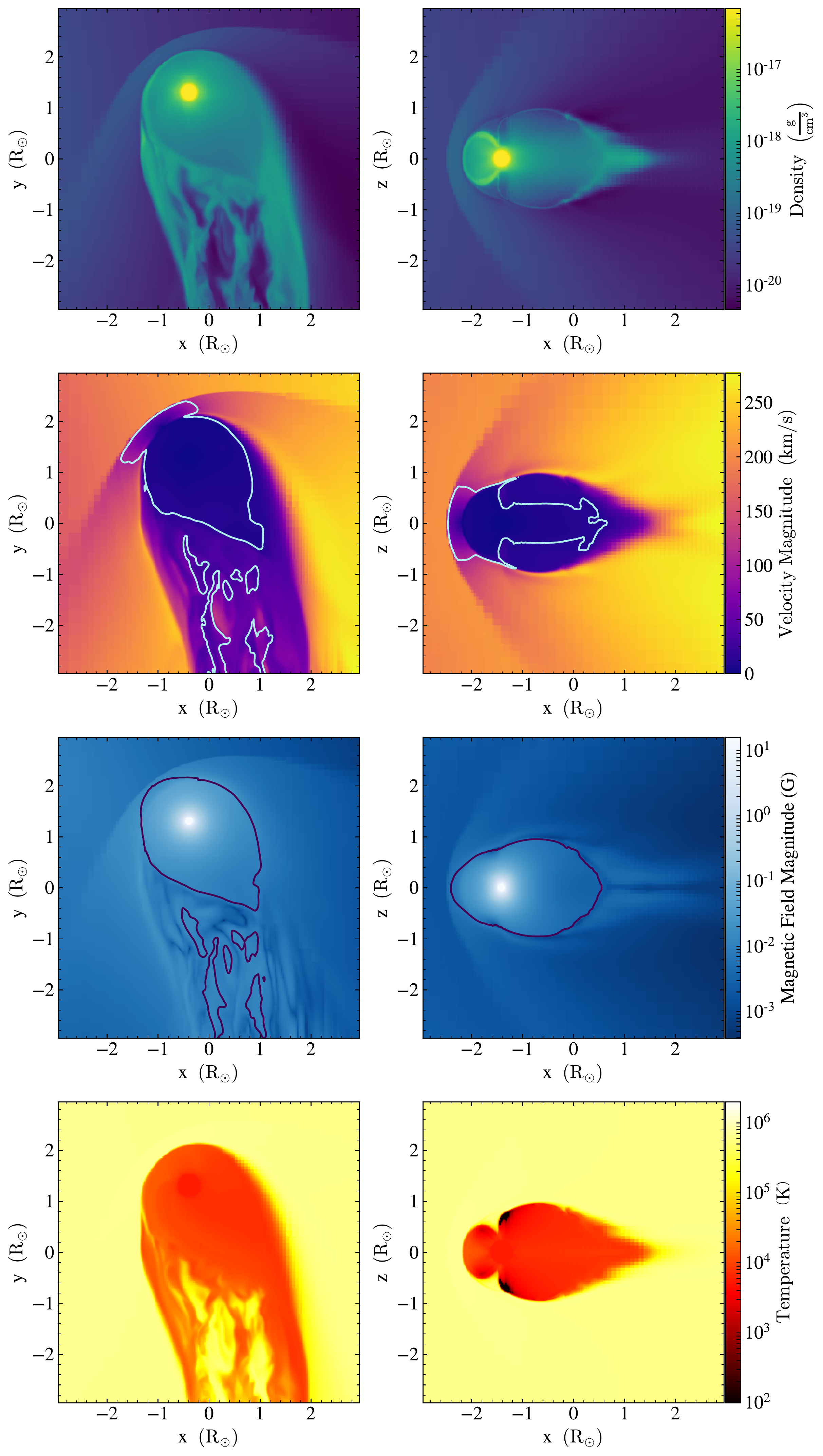}
\caption{
Density, velocity magnitude (with sonic surface contour), magnetic field magnitude (with Alfv\'{e}nic Mach surface contour) and temperature plots of the magnetosphere. Left column: top down view showing the planet, bow shock and start of cometary tail. Right column: same as left column but for a plane parallel to the z-axis. The plane intersects the centre of the planet and the apex of the bow shock. Mass-loss from the planet has filled the magnetosphere with material from the planet's atmosphere, confined by the planetary magnetic field. This material overflow from the downstream part of the magnetosphere to form the cometary tail.
\label{fig:magnetosphere}}
\end{figure*}

The bow shock and thin cometary tail is shown in greater detail in Fig. \ref{fig:magnetosphere} where (from top row to bottom) density, velocity magnitude, magnetic field magnitude and temperature are plotted for both a top down (left column) and side on (right column) views. Each of these quantities will be detailed in the following sections. 

\begin{figure}
\centering
\includegraphics[width=0.49\textwidth]{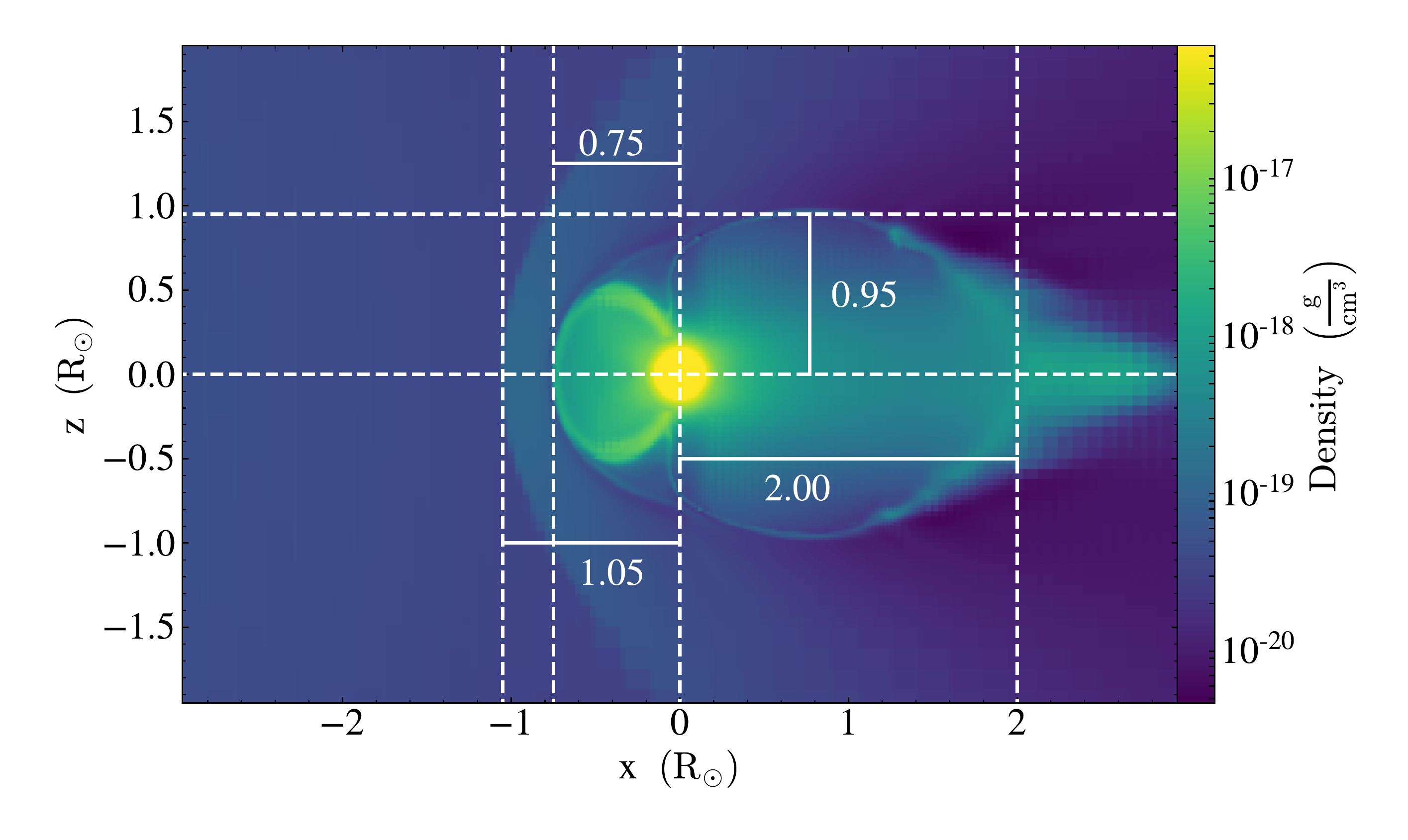}
\caption{
Diagram illustrating the position and extent of the separate parts of the magnetosphere. Dotted lines highlight the planets center as well as the edges of the magnetosphere. Solid lines show the width of each feature. The total length of the magnetosphere is given by $D_{\mathrm{M}} = 0.75 + 2 = 2.75 \ \mathrm{R_{\ast}}$. The magnetospheric radius, used in equation (\ref{eq:alpha}) to determine the frequency of ECMI emission is $R_{\mathrm{M}} = 0.75 \ \mathrm{R_{\ast}}$. The effective radius seen by the stellar wind and used in equation \ref{eq:radio_p} to calculate the power available to the ECMI process is $R_{\mathrm{eff}} = 0.95 \ \mathrm{R_{\ast}}$. The stand-off distance of the bow shock, $R_{\mathrm{SO}} = 1.05 \ \mathrm{R_{\ast}}$ is also shown.
}
\label{fig:Rm}
\end{figure}

\begin{figure}
\centering
\includegraphics[width=0.49\textwidth, trim={0 0 0 1cm},clip]{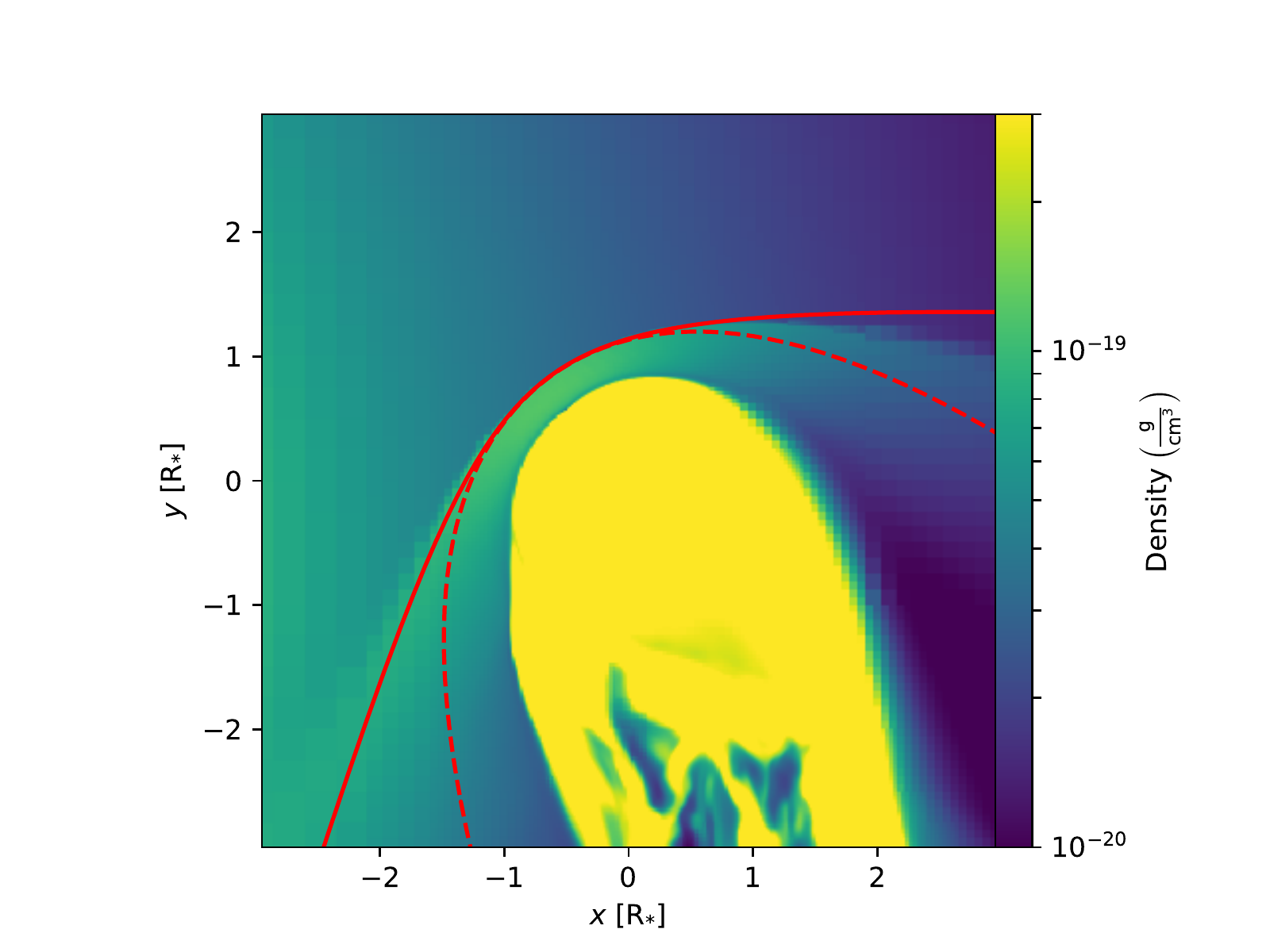}
\caption{
Comparison between predictions for the bow shock shapes. The solid red line indicates the prediction according to the work of \protect\cite{Tarango-Yong2018} and the dashed red line the prediction of \protect\cite{Wilkin1996}. The underlaying simulated bow shock can be seen as the over dense region upstream of the magnetosphere, with the colour map range chosen to highlight the shape of the bow shock for ease of comparison. Both simulated and analytic results agree well near the apex of the shock. Out to the sides the agreement begins to break down with both curves diverging from the simulation results. The prediction of \protect\cite{Tarango-Yong2018} is the closest to the simulation results.
}
\label{fig:bowshockanalytic}
\end{figure}

\subsubsection{Density Structure and Bow Shock}

The planetary wind has expanded into the magnetosphere up to the magnetopause where it accumulates until it reaches equilibrium and overflows, escaping from the aft part of the magnetosphere forming a cometary tail. Material following open field lines which protrude from both poles can escape from the magnetosphere and mix with the stellar wind. this happens in at the polar cusps and is most apparent in the right hand column.

For simplicity, \cite{Llama2013} states that the radius of the magnetosphere, $R_{\mathrm{M}}$, is also the distance from the exoplanets center to the bow shock apex. This is not strictly true as can be seen in Fig. \ref{fig:Rm}, which illustrates the various positions of the magnetospheric features relative to the exoplanets center and shows that the stand-off distance is $1.05 \ \mathrm{R_{\ast}}$ and $R_{\mathrm{M}} = 0.75 \ \mathrm{R_{\ast}}$ a value $\sim 30\%$ smaller. This difference between $R_{\mathrm{M}}$ and the stand-off distance is expected and it is observed for the Earth where the region between $R_{\mathrm{M}}$ and the stand-off distance is known as the magnetosheath. For the interest of clarity, in the following section, the distance to the apex will be referred to as the stand-off distance or $R_{\mathrm{SO}}$.

The relative velocity of the exoplanet to the oncoming stellar wind is supersonic, $|u_{\mathrm{K}} - u_{\phi}|~>~c_{\mathrm{s}}$, leading to the formation of a bow shock ahead of the exoplanet and it's magnetosphere. The angle made by the bow shock to the direction of orbital motion is given by $\theta_{0}~=~\arctan(u_{\mathrm{w}}/|u_{\mathrm{K}}~-~u_{\phi}|)$ \cite{Vidotto2010, Llama2013}, where $u_{\phi}$ is the azimuthal wind velocity at the orbial radius and $u_{\mathrm{K}}$ is the Kepler orbital velocity. For large orbital radii $\theta_{0}~\rightarrow~90^{\circ}$ (side on to the orbital motion), for small orbital radii $\theta_{0}~\rightarrow~0^{\circ}$ (directly in the path of the exoplanet). In practice $0^{\circ}~<~\theta_{0}~<~90^{\circ}$ and for this simulation $\theta_{0}~=~35.4^{\circ}$. \cite{Wilkin1996} derived an expression for the shape of the bow shock, giving the distance from the planets center to the shock $r_{\mathrm{shock}}$, as a function of $\theta_{0}$:
\begin{equation}
  r_{\mathrm{shock}} = \frac{R_{\mathrm{SO}}}{\sin(\theta - \theta_{0})} \left[ 3 \left( 1 - \frac{\theta - \theta_{0}}{\tan(\theta - \theta_{0})} \right) \right]^{\frac{1}{2}}.
  \label{eq:wilkinoid}
\end{equation}
The coordinate $\theta$ forms the angle between the orbital direction and the the apex of the shock. $u_{\mathrm{K}}$ is determed form $(G M_{\ast}/a)^{1/2}$ and $u_{\mathrm{w}}$ is sampled directly from the simulated stellar wind immediately a head of the planet. The stand-off distance of the bow shock is $1.05 \ R_{\mathrm{\ast}} = 6.97 \ R_{\mathrm{\circ}}$ which is taken from Fig. \ref{fig:Rm}, as the distance between the apex of the bow shock and the exoplanets center. Bow shocks described by equation (\ref{eq:wilkinoid}) have become known as \textit{wilkinoids} \citep{Cox2012, Tarango-Yong2018} and assume a plane parallel incident wind. For the case of a colliding wind they have become known as \textit{cantoids} \citep{Tarango-Yong2018}. 

Good agreement is found between the parabola described by equation (\ref{eq:wilkinoid}) and the simulation close to the apex of the bow shock as $\theta$ increases the agreement begins to break down. The wings of the simulated bow shock are supported further upstream compared to the prediction of equation (\ref{eq:wilkinoid}) which does not account for the presence of the magnetosphere, which leads to the conclusion that the bow shock in this work is not \textit{wilkinoid} in nature. The two curves and their deviation are shown in Fig. \ref{fig:bowshockanalytic}. To account for the over-supported wings of the bow shock in the \textit{wilkinoid} prediction, a model based on the work of \cite{Tarango-Yong2018} in which bow shock morphology is based upon a sophisticated treatment of parametric equations. Inputs to this model are dimensionless parameters allowing the model to be fitted to data or simulation results. These parameters are
\begin{equation}
  Q = - \frac{b^2}{a^2}, \ \Pi = \frac{aQ}{a-x_{0}}, \ \Lambda = \sqrt{Q \left( \frac{a + x_{0}}{a - x_{0}} \right)}.
  \label{eq:parametric}
\end{equation}
A by eye fit for this simulation are gives values of $Q = -4.6$, $\Pi = 2.7$ and $\Lambda = 1.4$. By substitution and algebraic manipulation of equations (\ref{eq:parametric}), $a$ and $b$ are found. The Cartesian coordinates of the curve are then given by
\begin{equation}
  x = x_{0} + a \cosh(t), \ y = b \sinh(t),
  \label{eq:bowcurve}
\end{equation}
with parametric variable $t \in [0, \pi]$. The interested reader is directed to \cite{Tarango-Yong2018} for a full description of the procedure. $r_{\mathrm{shock}}(\theta)$ and $\theta$ are recovered from Cartesian to polar coordinates via $\theta = \tan^{-1}(y/x)$ and $r_{\mathrm{shock}} = \sqrt{x^{2} + y^{2}}$. This model will be referred to as \textit{thoid} model (after the authors of the afore mentioned paper) for the remainder of the is paper. The \textit{thoid} curve is plotted in Fig. \ref{fig:bowshockanalytic} as the solid red line and more closely matches the wings of the shock than the \textit{wilkinoid} curve.

The \textit{thoid} model is derived based on the notion of a colliding wind system where the object around which the bow shock forms and the source of the wind in which it resides are point-like sources. This is precisely the situation that is found in the simulation presented here. A result that leads to the conclusion that the bow shock of a HJ needs to be modelled as a \textit{thoid} type bow shock.

Symmetry is often invoked in the characterization of bow shocks, however this symmetry is broken when the bodies around which the shocks form are in orbital motion \citep{Stevens1992, Gayley2009}. Such symmetry breaking is present in Fig. \ref{fig:bowshockanalytic}, with the left wing of the shock undergoing greater compression than the right. This is due to the relative differences in density and velocity either side of the planet. At the orbital radius, the velocity and density gradients are high relative to the extended wind, this is apparent in Fig. \ref{fig:parker_res}. This leads to the asymmetry seen in the figure and accounts for the deviation from the model bow shock described by equation (\ref{eq:bowcurve}). For HJs orbiting at larger radii, the gradients will be shallower leading to a more symmetric bow shock.

A proper treatment of the morphology of the bow shock would require an expression which gives the distance to the bow shock as $r_{\mathrm{shock}}(\theta) \rightarrow r_{\mathrm{shock}}(\theta, \phi)$ i.e. a three-dimensional representation which captures the non-spherical nature of the magnetosphere, an endeavour beyond the scope of this study.

\subsubsection{Velocity Field}

The bow shock, magnetosphere and tail are all apparent in the velocity maps. The bow shock exhibits a a distinctive jump corresponding to the jump in density seen in the panel above with the apex of the shock undergoing the greatest change in velocity. Within the magnetosphere, the velocity is approximately constant and $< 50 \ \mathrm{km/s}$, indicating either that the terminal velocity of the planetary wind is reached before the stand off distance of the magnetopause or that the magnetosphere is saturated with material which is then pressure supported leading to a corresponding reduction in velocity. The later conclusion is supported by inspection of the sonic surface position (see contour in the second row of Fig. \ref{fig:magnetosphere}), covering approximately the width of the magnetosphere in the equatorial plane. This is not the case for the initial conditions, where the sonic surface forms a sphere centred on the planet with a radius of $0.5 \ R_{\ast}$, about half the width of the steady state solution. As the simulation evolves the planetary wind expands into the magnetosphere while undergoing compression by the incident stellar wind, the magnetically confined material increases and with it the radius of the sonic surface, out to the magnetopause. 

large velocity gradients exist between the planetary and stellar winds of the order of $150 \ \mathrm{km/s}$, leading to a high degree of velocity shear and a Kelvin-Helmholtz (KH) unstable boundary. As the results show, there is no such instability present. A number of factors can act to inhibit this type of instability. A magnetic field parallel to the direction of the velocity field can suppress the instability if the Alfv\'{e}nic Mach number, $M_{\mathrm{Alf}}$, is of the order of unity in the shear layer \citep{Frank1996, Ryu2000}, the contour in the magnetic field plot (row three of Fig. \ref{fig:magnetosphere}), $M_{\mathrm{Alf}} = 1$ in the region of the velocity shear and thus suppresses KH instabilities. Non-magnetised HJs may exhibit KH behavior.

Numerical dissipation can also lead to a suppression of KH instabilities, Riemann solvers less conservative than the HLLC solver lead to diffusion at the velocity interface resulting in a smooth transition with no instability. The simulations presented here were conducted using algorithms known to be capable of producing KH behavior, leading to the conclusion that either the resolution employed is insufficient to resolve these features or that the magnetic suppression described above is responsible for the absence of the KH instability. 

Comparing the sonic surface and Alfv\'{e}nic Mach surface in the second and third rows of the right hand column of Fig. \ref{fig:magnetosphere}, one can see that the sonic surface almost reaches the poles of the planet, while the Alfv\'{e}nic Mach surface remains at the magnetopause and therefore material out of the equatorial plane is primarily supported by magnetic tension rather than ram pressure from the planetary wind.

\subsubsection{Magnetic Field Topology}

The planetary magnetic field remains largely dipolar apart from an antisymmetric perturbation aft and fore of the planet. The asymmetric shape of the magnetosphere is due to the ram pressure of the stellar wind on the field lines, compressing upstream and elongating downstream, where the planets magnetosphere is stretched out to form a magnetotail and plasma sheet. The 3D representation of the field lines in Fig. \ref{fig:volume} also shows this asymmetry. From the third row in Fig. \ref{fig:magnetosphere}, it can be seen that the magnetosphere downstream of the exoplanet is approximately twice the radius of the upstream magnetosphere. This makes estimating the magnetospheric radius, $R_{\mathrm{eff}}$, challenging as the closed field line region described in Section \ref{sec:emission} can no longer be guarantied to reflect the effective width of the magnetosphere, as seen by the oncoming stellar wind. As such, the $R_{\mathrm{eff}}$ will be estimated from the the extent magnetosphere cross section, this is shown in more detail in Fig. \ref{fig:Rm}, see Section \ref{sec:radio_p}.

As mentioned in the previous section, Fig. \ref{fig:magnetosphere} indicates the $M_{\mathrm{Alf}} = 1$ surface, within which the exoplanetary wind is sub Alfv\'{e}nic and perturbations to the magnetic field can travel back to the exoplanets surface. Outside this surface the wind is super Alfv\'{e}nic meaning that the velocity is faster than the speed at which magnetic perturbations travel and is undisturbed by activity outside this surface unless the flow undergoes rapid change. This can be the result of, for example a Coronal Mass Ejection (CME). The present simulation is conducted with a steady stellar wind, so this form of perturbation is absent. For a sufficiently active stellar host, which exhibits CME or other forms of rotational or time dependent events, the stability of the magnetosphere of a HJ will be come a time dependent problem. 

\subsubsection{Temperature Distribution}

The temperature profiles vividly illustrate the extent to which the exoplanetary wind expands from the surface. In both the left and right hand columns, the ambient stellar wind is constant at the stellar surface temperature apart from the exoplanetary material confined to the magnetosphere and the cometary tail. The bow shock is not apparent in the temperature maps and the polar cusps are the same temperature as the stellar wind. This indicates that the stellar wind has penetrated the magnetosphere and can provide a supply of energetic electrons to the poles of the HJ for the ECMI process.

Another feature exhibited by both the northern and southern polar cusps is two dramatically lower temperature regions on the downstream side of the magnetosphere, these can bee sen as the black regions in the bottom right plot of Fig. \ref{fig:magnetosphere}. The temperature in these region is $\sim 100 \ \mathrm{K}$ despite being directly adjacent to the stellar wind ($10^{6} \ \mathrm{K}$). The presence of these features may indicate the limit of the isothermal assumptions made in the construction of the model described in Section \ref{sec:model_init}. The inclusion of thermal conduction, cooling and a non-unity $\gamma$ in equation (\ref{eq:energy}) would be necessary to test this limit. It should be noted that the lack of temperature gradients within the magnetosphere is in agreement with the isothermal nature of the simulations. Together with this and in the absence of a comparable non-isothermal study, it is assumed that these features are of negligible importance to the structure of the planets magnetosphere. 

\subsection{Radio Power and Frequency}
\label{sec:radio_p}

\begin{figure}
\centering
\includegraphics[width=0.49\textwidth,trim={8cm 0 8cm 0},clip]{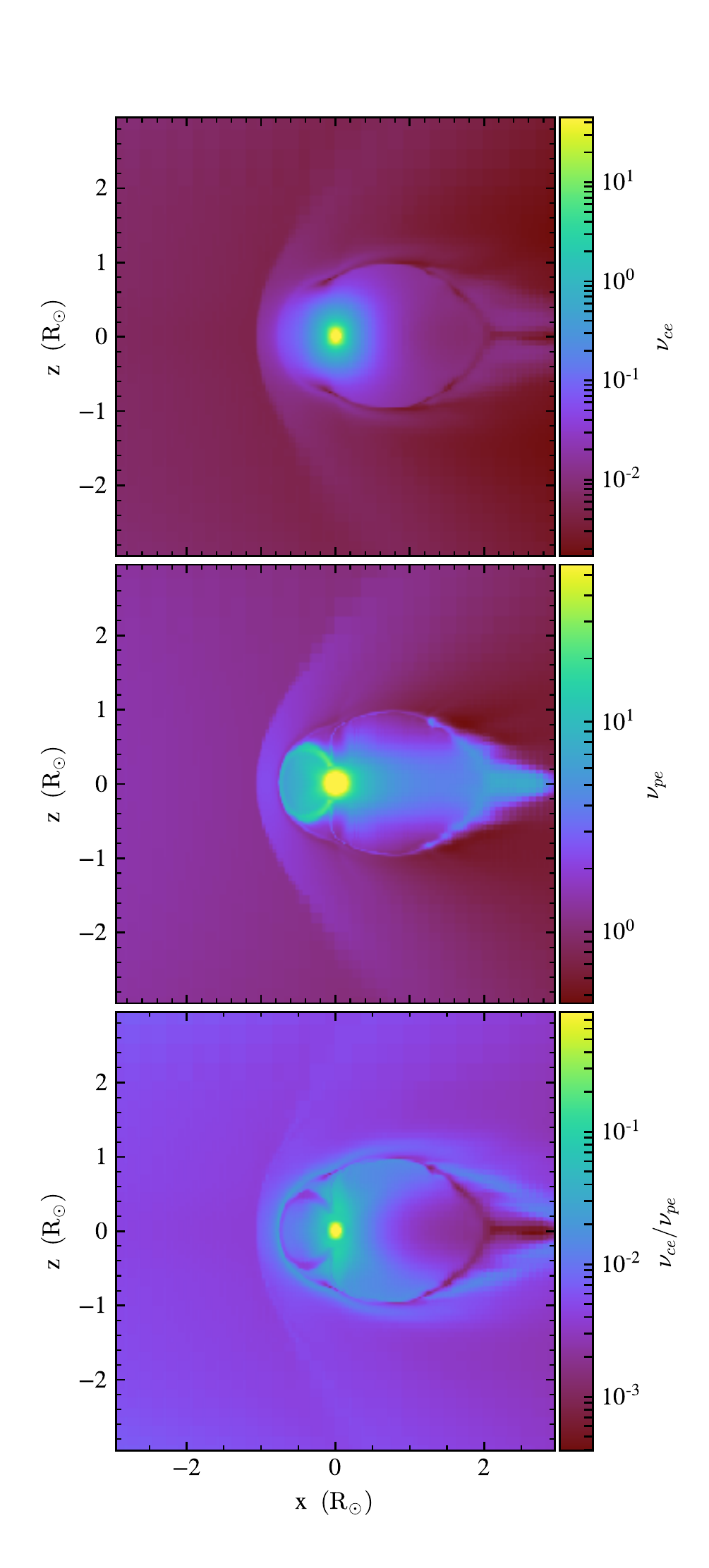}
\caption{
The immediate vicinity of the exoplanet with $\nu_{\mathrm{ce}}$ (top), $\nu_{\mathrm{pe}}$ (middle) and $\nu_{\mathrm{ce}} / \nu_{\mathrm{pe}}$ (Bottom). Features such as the bow shock, radiation belts and magnetopause are visible in each. The maximum of $\nu_{\mathrm{ce}} / \nu_{\mathrm{pe}} \sim 0.1$, an order of magnitude below the value necessary for the ECMI process to lead to radio cyclotron emission. This value is also $\times 25$ lower than the efficiency criterion stated in Section \ref{sec:ECMI}. The expanding atmosphere of the exoplanet raises the plasma frequency in the magnetosphere, lowering the ratio $\nu_{\mathrm{ce}} / \nu_{\mathrm{pe}}$ and inhibiting the ECMI process.
\label{fig:fcfp}}
\end{figure}

\begin{figure*}
\centering
\includegraphics[width=0.99\textwidth, trim={0 2cm 0 2cm},clip]{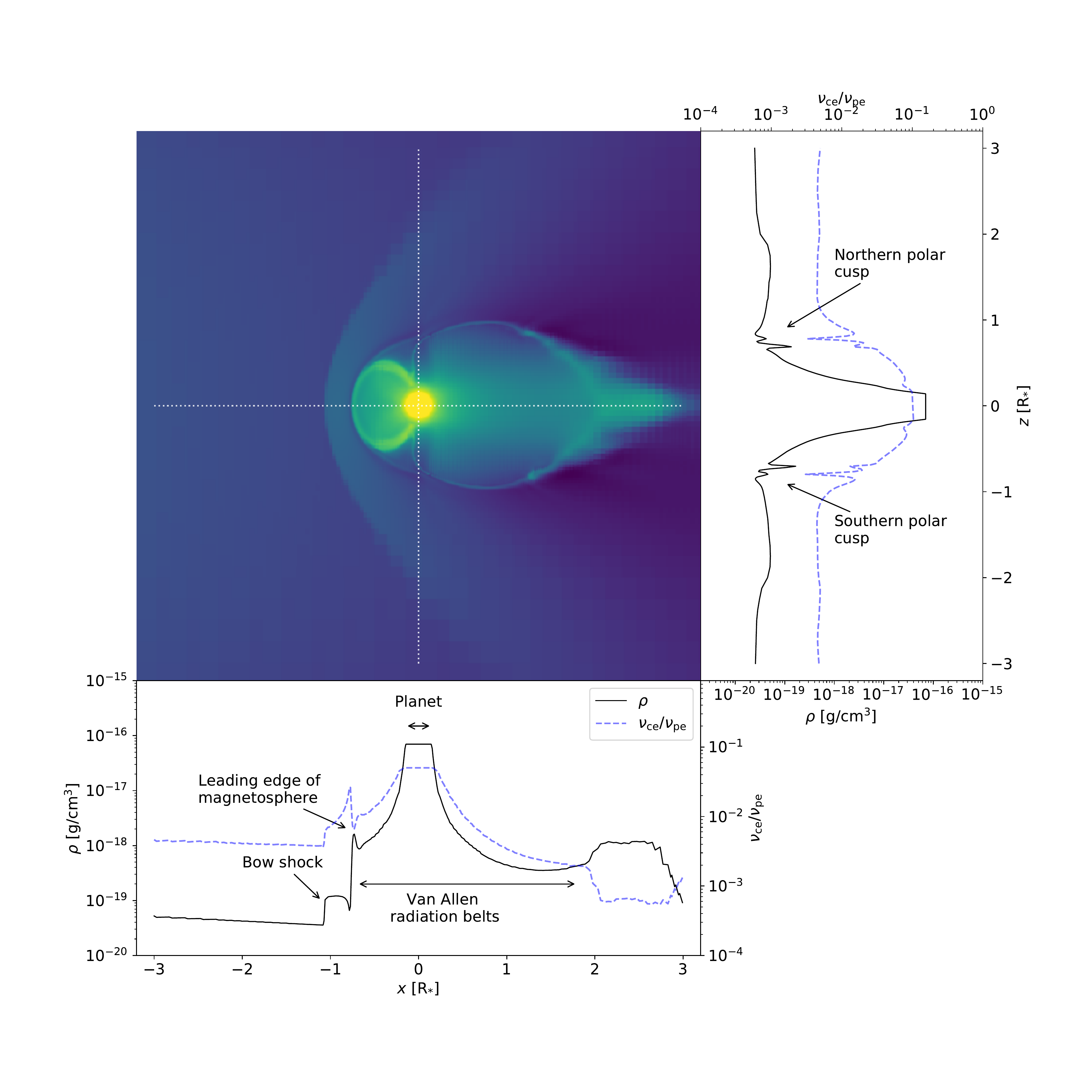}
\caption{
Top left: Density slice plot showing the exoplanet, magnetosphere and bow shock. The dotted white lines indicate the sampling used for the profiles shown below and to the right. Bellow: $\rho$ and $\nu_{\mathrm{ce}} / \nu_{\mathrm{pe}}$ profiles for the sample line parallel to the x-axis indicating the location of the planet, bow shock, leading edge of the magnetosphere and the equivalent Van Allen radiation belts. Right: Same as below but for the sample line parallel to the z-axis. Indicated are the the northern and southern polar cusps which form entry points to the planets magnetosphere for stellar electrons.
\label{fig:profile_fcfp}}
\end{figure*}

A determining factor in the intensity, and therefore, detectability of radio emission from HJs is the power available to the ECMI process. Incident kinetic and magnetic energy from the stellar wind on the magnetosphere is translated to radio power. Section \ref{sec:emission} describes this process, with equation (\ref{eq:radio_p}) giving the theoretical radio power.

Fig. \ref{fig:Rm} highlights the separate parts of the magnetosphere including the effective radius, $R_{\mathrm{eff}}~=~0.95 \ R_{\ast}$. This value together with the stellar mass-loss rate $\dot{M}_{\ast}~=~2.14 \times 10^{12} \ \mathrm{g/s}$, stellar wind velocity as seen by the planet, $u_{\mathrm{w}}~=~207 \ \mathrm{km/s}$ and the parameters from Table \ref{tab:params}, equation (\ref{eq:radio_p}) is used to determine a radio power $P_{\mathrm{r}}~=~1.42~\times~10^{19} \ \mathrm{erg/s}$ available to the ECMI. Assuming a typical distance of $10 \ \mathrm{pc}$ between the exoplanet system and the observer, the resultant radio flux is $0.069 \ \mathrm{mJy}$. 

This flux is emitted at a range of frequencies between $\nu(B(\alpha))$, as described in Section \ref{sec:emission}, and $\nu_{\mathrm{peak}}$. The oncoming stellar wind has compressed the magnetosphere to such a degree that determining $\alpha$ via the transition from open to closed to field lines is not possible. The approximation, equation (\ref{eq:alpha}), is used with $R_{\mathrm{M}} = 0.75$ from Fig. \ref{fig:Rm} instead to give $\alpha~=~23.5^{\circ}$ and $B(\alpha)~=~B_{\mathrm{eq}} \left( 1 + 3 \cos^2(\alpha) \right)^{1/2}~=~1.38 \ \mathrm{G}$ and therefore $\nu(B(\alpha))~=~3.88 \ \mathrm{MHz}$ which will serve as a lower bound for $\nu_{\mathrm{ce}}$. The upper limit, $\nu_{\mathrm{peak}}$, is assumed to be a function of the polar field strength, $B_{\mathrm{pole}} = 2 B_{\mathrm{eq}} = 2 \ \mathrm{G}$ and therefore, using equation (\ref{eq:ce}), $\nu_{\mathrm{peak}} = 5.60 \ \mathrm{MHz}$. This leads to the range $3.88 \ \mathrm{MHz}~<~\nu_{\mathrm{ce}}~<~5.60 \ \mathrm{MHz}$ for the ECMI radio emission. As stated in Section \ref{sec:detect} the ionospheric cut off is 1-10~MHz, placing $\nu_{\mathrm{ce}}$ at best in the lower end of what is detectable and at worst below the detectable threshold. A result which alone makes detecting radio emission from HJs with magnetic fields of the order used in this study, a challenge. However, this range will be different for each HJ as it is a direct function of the magnetic field strength. Field strengths of $B_{\mathrm{eq}}~\sim~5 \ \mathrm{G}$ will have an upper limit of 28 MHz, placing it above the 10 MHz cutoff.

\subsection{Cyclotron Emission}

The frequency calculated in the previous section does not necessarily lead to detectable emission. The ECMI process is dependent upon two factors, as discussed in Section \ref{sec:ECMI},  the conditions must favour both a high $\nu_{\mathrm{ce}}$ and a low $\nu_{\mathrm{pe}}$. These two frequencies are evaluated for the region around the planet and plotted in Fig. \ref{fig:fcfp}. In this figure, both $\nu_{\mathrm{ce}}$ (top), $\nu_{\mathrm{pe}}$ (middle) are plotted along with their ratio $\nu_{\mathrm{ce}} / \nu_{\mathrm{pe}}$ (bottom). From equation (\ref{eq:ECMI}), the ratio needs to be greater than $\sim 2.5$ for the ECMI process to be efficient at generating radio emission. As can be seen in the bottom plot of Fig. \ref{fig:fcfp}, this condition is not met at any point within the vicinity of the HJ, with the largest ratio of $\nu_{\mathrm{ce}} / \nu_{\mathrm{pe}} \sim 0.1$ in the polar region. This value is $\times 25$ lower than the efficiency criterion stated in Section \ref{sec:ECMI}. The expanding atmosphere of the planet has raised $\nu_{\mathrm{pe}}$ to such a degree that it inhibits any emission from electrons undergoing the ECMI process. With the implication that such an exoplanet as the one simulated here, the ECMI process is inefficient at generating radio emission. 

This result is in agreement with \cite{Weber2017} who, through analytic modelling, found very similar results for a range of exoplanet parameters including magnetic field strengths of the order $50 \ \mathrm{G}$. With the conclusion that, for efficient emission generation, a HJs magnetic field would need to be $> 50 \ \mathrm{G}$. Recent work by \cite{Yadav2017} has investigated the energy available to the planetary dynamo in the form of absorbed radiation from HJ hosts. Their work found that the majority of HJs should have magnetic fields in the range $50 \ \mathrm{G}~<~B_{\mathrm{pole}}~<~150 \ \mathrm{G}$ (see Fig. 2(b) of afore mentioned paper). However due to the work of \cite{Weber2017}, one would conclude that such HJs should have been detected, casting doubt on the high magnetic field strength model. Other factors also play a role in the  much also be considered such as; the intermittent nature of emission, beaming direction and the sparse number of observations. All factors which play a role when considering the lack of radio detections of HJs.

Early resutls of the simulations presented here were communicated in \cite{Daley-Yates2017}, in which two regions above and below the HJ poles were found to have $\nu_{\mathrm{ce}} / \nu_{\mathrm{pe}}~>~1$. These simulations have been refined and now use a more sophisticated treatment of the magnetic field, handling the low plasma-$\beta$ environment near the HJs surface resulting in a more stable simulation. As such this work builds upon these simulation.

To further investigate the relation between flow features, the HJ's magnetosphere and the ECMI efficiency ratio $\nu_{\mathrm{ce}} / \nu_{\mathrm{pe}}$, Fig. \ref{fig:profile_fcfp} displays a density slice through the HJ, showing both poles, bow shock and extended magnetosphere. Two lines intersecting at the planets center indicating the sample lines of the plots shown to the right and bottom of the figure. $\nu_{\mathrm{ce}} / \nu_{\mathrm{pe}}$ is sensitive to all features present in the density structure. In the case of the $x$-axis aligned profile, the bow shock, edge of the magnetosphere and the planet itself give rise to increases in $\nu_{\mathrm{ce}} / \nu_{\mathrm{pe}}$. For the $z$-axis aligned profile, the main features are the polar cusps with $\nu_{\mathrm{ce}} / \nu_{\mathrm{pe}}$ fluctuating across them. In both profiles, the highest ratio is found at the planets surface, in agreement with the previous section.

\section{Conclusions}

Radio emission due to the ECMI process in HJ exoplanets is expected to be considerable, however, to the contrary of numerous theoretical works, no repeatable detections have been made to date. This work provides an explanation for this through the use of MHD simulations. Global evolution and the circumplanetary environment of a HJ hosting system has been investigated via rigorous treatment of both the stellar and planetary winds and magnetic field in order to determine the efficiency of the ECMI process for producing detectable radio emission.

The frequency of emission has been calculated using the the model of \cite{Vidotto2011} as a lower limit for the emission and the polar magnetic field strength as an upper limit. This gives the range $3.88 \ \mathrm{MHz}~<~\nu_{\mathrm{ce}}~<~5.60 \ \mathrm{MHz}$ for the emission frequency. this is below or close to the Earth's ionospheric cutoff frequency and detection limits of current instruments. HJs with field strength greater that used here may result in emission frequencies above this cut-off point.

For the simulated HJ, the ECMI process is completely inhibited by the expanding atmosphere, due to the UV radiation form the host star. This result is in close agreement with the analytic work of \cite{Weber2017}. For such an exoplanet to produce detectable emission, the magnetic field is required to be considerably greater than that used here. As no such detections exist, credence is given to a weak magnetic field model which is used in this work and powered by slow rotation at the orbital frequency (tidal locking), over the strong magnetic field, radiation powered model of \cite{Yadav2017}.

Analysis of the bow shock has also been conducted showing that the model of \cite{Tarango-Yong2018} (\textit{thoid}) better recreates the shock produced by a HJ than the more traditional \cite{Wilkin1996} (\textit{wilkinoid}) model.

Results presented in this work do not rule out the possibility for radio frequency detection of HJs which experience energetic or transient events such as CME which act to compress and deform the exoplanetary magnetosphere. These transient events are left for future study.

\section{Acknowledgments}

The authors thank the anonymous reviewer for their helpful comments and suggestions; which improved the quality and content of the publication.

The authors acknowledge support from the Science and Technologies Facilities Research Council (STFC). 

Computations were performed using the University of Birmingham's BlueBEAR HPS service, which was purchased through HEFCE SRIF-3 funds. See http://www.bear.bham.ac.uk.

\bibliographystyle{mnras}

\bibliography{/Users/simon/Work/Reading/References/ExoplanetRefs/exoplanet_refs,/Users/simon/Work/Reading/References/MHDRefs/mhd_refs,/Users/simon/Work/Reading/References/CollidingWindsRefs/colliding_wind_refs,/Users/simon/Work/Reading/References/EmissionMechanisms/emission_mechanisms}

\label{lastpage}

\end{document}